\documentclass[fleqn,10pt]{wlscirep}

\usepackage[english]{babel}
\usepackage{graphicx} 
\usepackage[title]{appendix}

\title{The noisy voter model on complex networks}

\author[1,*]{Adri\'an Carro}
\author[1]{Ra\'ul Toral}
\author[1]{Maxi {San Miguel}}
\affil[1]{IFISC (CSIC-UIB), Instituto de F\'isica Interdisciplinar y Sistemas Complejos,
Campus Universitat de les Illes Balears, E-07122, Palma de Mallorca, Spain}

\affil[*]{adrian.carro@ifisc.uib-csic.es}

\begin{abstract}
We propose a new analytical method to study stochastic, binary-state models on complex networks. Moving beyond the usual mean-field theories, this alternative approach is based on the introduction of an annealed approximation for uncorrelated networks, allowing to deal with the network structure as parametric heterogeneity. As an illustration, we study the noisy voter model, a modification of the original voter model including random changes of state. The proposed method is able to unfold the dependence of the model not only on the mean degree (the mean-field prediction) but also on more complex averages over the degree distribution. In particular, we find that the degree heterogeneity ---variance of the underlying degree distribution--- has a strong influence on the location of the critical point of a noise-induced, finite-size transition occurring in the model, on the local ordering of the system, and on the functional form of its temporal correlations. Finally, we show how this latter point opens the possibility of inferring the degree heterogeneity of the underlying network by observing only the aggregate behavior of the system as a whole, an issue of interest for systems where only macroscopic, population level variables can be measured.
\end{abstract}

\begin{document}

\flushbottom
\maketitle
%
%
\thispagestyle{empty}




\section{Introduction}


Stochastic, binary-state models have been used to study the emergence of collective phenomena in a wide variety of systems and fields. Examples range from classical problems in statistical physics, such as equilibrium and non-equilibrium phase transitions \cite{Gunton1983,Marro1999}, to biological and ecological questions, such as neural activity \cite{Hopfield1982} and species competition \cite{Clifford1973,Crawley1987}, or even to social and epidemiological topics, such as the spreading of diseases in a population \cite{Anderson1992,PastorSatorras2001,Watts2002,Serrano2006,Castellano2009,Castellano2012}. In general, these systems are considered to be embedded in a network structure, where the nodes are endowed with a binary-state variable ---spin up or down--- and the links between nodes represent the interactions or relations between them. While most of these models were initially studied in regular lattices, there has recently been a growing interest in more complex and heterogeneous topologies \cite{Barabasi2002,Newman2003,Barrat2008,Newman2010}. An important result of these recent works has been to show that, for a given model, the structure of the underlying network may strongly influence the dynamics of the system and affect its critical behavior, leading, for instance, to different critical values of the model parameters \cite{Lambiotte2007,Gleeson2011,Vilone2012,Gleeson2013}. This has been shown to be the case, for example, for the critical temperature of the Ising model \cite{Dorogovtsev2002,Leone2002,VianaLopes2004}, for the epidemic threshold in spreading phenomena \cite{Boguna2003,Durrett2010,Castellano2010,Parshani2010}, and for the mean return and first-passage times in random walks \cite{Masuda2004,Sood2005}. Thus, the quantification of the effect of the underlying topology on such systems and dynamics is, from a practical point of view, a matter of prime importance.


A paradigmatic example of this kind of models, with applications in the study of non-equilibrium systems in a wide range of fields, is the voter model \cite{Clifford1973,Holley1975}. Based solely on local, pairwise interactions, this model assumes that, in a single event, a randomly chosen node copies the state of one of its neighbors, also chosen at random. In this paper, however, we are going to focus on the noisy voter model, a variant of the original voter model which, apart from pairwise interactions, includes random changes of state. This variant model has been studied by, at least, four mutually independent strands of research, largely unaware of each other and belonging to different fields. Namely, percolation processes in strongly correlated systems \cite{Lebowitz1986}, heterogeneous catalytic chemical reactions \cite{Fichthorn1989,Considine1989}, herding behavior in financial markets \cite{Kirman1993}, and probability theory \cite{Granovsky1995}. While both the first and the last strands of literature are directly inspired by the voter model, explicitly using terms such as ``noisy voter model'' or similar, contributions in the contexts of catalytic reactions and financial markets do not refer to the voter model, and use terms such as ``catalytic reaction'', ``herding'' or ``Kirman model'' instead. More recently, the inclusion of random events in the voter model has also been used to reproduce some of the statistical regularities observed in real electoral processes \cite{FernandezGracia2014}.


For any finite system, the behavior of the noisy voter model is characterized by the competition between two opposing mechanisms, related to two different types of noise. On the one hand, the pairwise interaction mechanism is related to interfacial fluctuations (internal noise) and tends to order the system, driving it towards a homogeneous configuration ---all spins in the same state, whether up or down---. Depending on the dimension of the system, this mechanism leads to a coarsening process or to a metastable partially ordered state, both of them perturbed by finite-size fluctuations (one of which eventually drives the system to full order). In the absence of any other mechanism, as it is the case in the original voter model, the homogeneous configurations become absorbing states of the dynamics \cite{AlHammal2005}. On the other hand, the random change mechanism is related to thermal-like fluctuations (external noise) and tends to disorder the system, pulling it from the homogeneous configurations. Therefore, this second mechanism leads to the disappearance of the typical absorbing states of the voter model and to the restoration of ergodicity \cite{Granovsky1995}. The main consequence of this competition is the appearance of a noise-induced, finite-size transition between two different behavioral regimes ---a mostly ordered regime dominated by pairwise interactions and a mostly disordered regime dominated by noise \cite{Kirman1993,Alfarano2008}. While the effect of different network topologies on the behavior of the voter model has been well established \cite{Suchecki2005,Sood2005B,Suchecki2005B,Vazquez2008}, the case of the noisy voter model has received much less attention, most of the corresponding literature focusing only on regular lattices \cite{Lebowitz1986,Granovsky1995} or on a fully-connected network \cite{Kirman1993,Alfarano2008}. Finally, the use of a mean-field approach in some recent studies considering more complex topologies \cite{Alfarano2009,Alfarano2011,Diakonova2015} did not allow to find any effect of the network properties ---apart from its size and mean degree--- on the results of the model.


In this paper, we move beyond the usual mean-field approximations \cite{Vazquez2008B,Alfarano2009,Diakonova2015} and propose an alternative analytical approach, based on an annealed approximation for uncorrelated networks and inspired by a recently introduced method to deal with heterogeneity in stochastic interacting particle systems \cite{Lafuerza2013}. In particular, we approximate the network by a complementary, weighted, fully-connected network whose weights are given by the probabilities of the corresponding nodes being connected in uncorrelated networks of the configuration ensemble \cite{Newman2003B,Boguna2004,Bianconi2009,Sonnenschein2012} ---i.e., proportional to the product of their degrees. Furthermore, we present a formulation of the problem in terms of a master equation for the probability distribution of the individual states of the nodes. In this way, we are able to find approximate analytical expressions for the critical point of the transition, for a local order parameter and for the temporal correlations. As opposed to previous mean-field approaches, we find that the degree heterogeneity ---variance of the underlying degree distribution--- has a significant impact on all of these variables, leading to a larger value of the critical point, a higher level of order and a modification of the functional form of the temporal correlations. As we will show, this latter point opens the possibility of inferring the degree heterogeneity of the underlying network by observing only the aggregate behavior of the system as a whole, an issue of interest for systems where only macroscopic, population level variables can be measured. Finally, these results are confirmed by numerical simulations on different types of networks, allowing for a constant mean degree ($\overline{k}=8$) while leading to different degree distributions. In particular, in order of increasing degree heterogeneity, we focus on Erd\"os-R\'enyi random networks \cite{Erdos1960}, Barab\'asi-Albert scale-free networks \cite{Barabasi1999} and dichotomous networks \cite{Lambiotte2007} (whose nodes are assigned one out of two possible degrees, in our case \mbox{$k_1 = \overline{k}/2$} or \mbox{$k_2 = \sqrt{N}$}).


\section{Model and methods}


\subsection{Model definition}


Consider a system composed of $N$ nodes in a given network of interactions. At any point in time, each node $i$ is considered to be in one of two possible states, and is therefore characterized by a binary variable \mbox{$s_i = \{0,1\}$}. Moreover, due to the network structure, each node $i$ is also characterized by a certain set of (nearest) neighbors, $nn(i)$, and by its corresponding degree or number of those neighbors, $k_i$. The evolution of the state of each node, $s_i$, occurs stochastically with probabilities that depend on the state of the updating node and on the states of its neighbors. In particular, these probabilities consist of two terms: on the one hand, there are random pairwise interactions between node $i$ and one of its neighbors \mbox{$j \in nn(i)$}, after which $i$ copies the state of $j$; and, on the other hand, there are random changes of state, playing the role of a noise. The transition rates for each node $i$ can be written as
    \begin{equation}
    \begin{aligned}
    r^+_i \equiv r \left( s_i=0 \rightarrow s_i=1 \right) &= a + \frac{h}{k_i} \sum_{j \in nn(i)} s_j \, ,\\[5pt]
    r^-_i \equiv r \left( s_i=1 \rightarrow s_i=0 \right) &= a + \frac{h}{k_i} \sum_{j \in nn(i)} (1 - s_j) \, ,
    \end{aligned}
    \label{e:rates}
    \end{equation}
where the noise parameter $a$ regulates the rate at which random changes of state take place, and the interaction parameter $h$ does so with the interaction-driven changes of state. Defined in this way, the noisy voter model becomes the network-embedded equivalent of the Kirman model in its original, extensive formulation \cite{Kirman1993,Alfarano2008,Alfarano2009}. Furthermore, note that, in the limit case of $a = 0$ and with an appropriate time rescaling, we recover the transition rates of the original voter model.



Let us stress here that, even if the model appears to have two parameters, one of them can always be used as a rescaling of the time variable, so that there is only one relevant parameter: the ratio between the two introduced coefficients, $a/h$. Indeed, only one parameter is introduced in the previous literature in the context of the noisy voter model \cite{Lebowitz1986,Granovsky1995,Diakonova2015}. On the contrary, prior works dealing with the Kirman model usually keep both parameters \cite{Kirman1993,Alfarano2008,Alfarano2009}. For consistency with one and the other strands of literature, we are going to consider both parameters explicitly in our analytical approach, while we keep the interaction parameter fixed as $h=1$ for our numerical results ---allowing the noise parameter $a$ to vary.

In order to characterize the global state of the system we introduce the global variable $n$, defined as the total number of nodes in state $s_i=1$,
\begin{equation}
    n = \sum_{i=1}^N s_i \, ,
    \label{e:n_definition}
\end{equation}
and taking values \mbox{$n \in {0,1,...,N}$}. Note that this variable does not take into account any aspect of the network structure. 


It should be observed that, for $a \neq 0$, there are no absorbing states in the model ---the probability to move from one state to any other is strictly positive--- and therefore the Markov chain is said to be ergodic: in the steady state, averages over time are equivalent to ensemble averages. In practice, the smaller $a$ is, the longer the time needed for both statistics to be actually equivalent. Thus, in the limit case of $a = 0$ the time needed becomes infinite, and we recover the voter model behavior: non-ergodicity with two absorbing states, at $n=0$ and $n=N$. Moreover, we are going to use the notation $\langle x \rangle$ for ensemble averages with random initial conditions, while we leave $\overline{f(k)}$ for averages over the degree distribution, i.e.,
\begin{equation}
    \overline{f(k)} = \frac{1}{N} \sum_{i=1}^N f(k_i) \, .
    \label{e:overbar}
\end{equation}
Similarly, we will differentiate between the variance of a variable $x$ over realizations, noted as $\sigma^2[x]$, and the variance of the degree distribution, labeled as $\sigma^2_k$. Note, nonetheless, that for the numerical steady state values to be presented in the following sections, averages are performed both over time and over an ensemble of realizations with random initial conditions, assuming an initial transient of $N$ time units.


\subsection{General formulation}


The stochastic evolution of the system can be formalized as a Markov process. In particular, we can write a general master equation for the \mbox{$N$-node} probability distribution \mbox{$P(s_1,\ldots,s_N)$} (see Supplementary Information) and use it to derive general equations for the time evolution of the first-order moments and the second-order cross-moments of the individual nodes' state variables $s_i$,
\begin{align}
    &\frac{d \langle s_i \rangle}{dt} = \langle r_i^+ \rangle - \langle (r_i^+ + r_i^-) s_i \rangle \, ,\label{e:first_moment}\\[5pt]
    &\frac{d \langle s_i s_j \rangle}{dt} = \;\langle r_i^+ s_j \rangle + \langle r_j^+ s_i \rangle - \langle q_{ij} s_i s_j \rangle + \delta_{ij} \left[ \langle s_i r_i^- \rangle + \langle (1 - s_i) r_i^+ \rangle \right] \, ,\label{e:second_moment}
\end{align}
where \mbox{$q_{ij} = r_i^+ + r_i^- + r_j^+ + r_j^-$} and $\delta$ stands for the Kronecker delta (see Appendices~\ref{a:equation_for_the_time_evolution_of_the_first-order_moments} and ~\ref{a:equation_for_the_time_evolution_of_the_second-order_cross-moments} for details). In general, if the transition rates depend on the individual state variables $s_i$, these equations involve higher order moments and they cannot be solved without a suitable approximation \cite{Lafuerza2013}. However, for the transition rates of the noisy voter model, due to their particular form, both equations become independent of higher order moments.


For the first-order moments, introducing the transition rates~\eqref{e:rates} into equation~\eqref{e:first_moment}, we obtain
\begin{equation}
    \frac{d \langle s_i \rangle}{dt} = a - (2a + h) \langle s_i \rangle + \frac{h}{k_i} \sum_{m \in nn(i)} \langle s_m \rangle \, ,
    \label{e:kirman_first_moment}
\end{equation}
an equation directly solvable in the steady state, when the influence of the initial conditions has completely vanished and thus $\langle s_i \rangle_{st}$ is independent of $i$. In this way, we find, for the steady state average individual variables $s_i$ and, by definition, for the steady state average global variable $n$, respectively,
\begin{equation}
    \langle s_i \rangle_{st} = \frac{1}{2} \, , \qquad \quad \langle n \rangle_{st} = \frac{N}{2} \, ,
    \label{e:steady_state_first_moment}
\end{equation}
the expected results given the symmetry of the system.


In the case of the second-order cross-moments, when we introduce the transition rates~\eqref{e:rates} into equation~\eqref{e:second_moment}, we obtain
\begin{equation}
\begin{aligned}
    \frac{d \langle s_i s_j \rangle}{dt} =& a (\langle s_i \rangle + \langle s_j \rangle) - 2(2a + h) \left\langle s_i s_j \right\rangle + \frac{h}{k_i} \sum_{m \in nn(i)} \langle s_m s_j \rangle + \frac{h}{k_j} \sum_{m \in nn(j)} \langle s_m s_i \rangle\\
    &+ \delta_{ij} \left[ a + h \langle s_i \rangle + \frac{h}{k_i} \sum_{m \in nn(i)} \langle s_m \rangle - \frac{2h}{k_i} \sum_{m \in nn(i)} \langle s_m s_i \rangle \right] \, ,
    \label{e:kirman_second_moment}
\end{aligned}
\end{equation}
which, even if independent of higher order moments, cannot be solved in the absence of an explicit knowledge of the network connections ---the adjacency matrix. This is due to the presence of sums over neighbors \mbox{$\sum_{m \in nn(i)}$} where the terms are not independent of the particular pair of nodes \mbox{$m,i$}. In order to find the corresponding steady state solution, we introduce below an approximation of the network allowing us to write the previous equation in terms of sums over the whole system.


\subsection{Annealed approximation for uncorrelated networks}


Given a complex network with adjacency matrix $A_{ij}$ and degree sequence \mbox{$\{k_i\}$}, we can use an annealed graph approach \cite{Vilone2004,Dorogovtsev2008,Guerra2010} to define a complementary, weighted, fully-connected network with a new adjacency matrix $\tilde{A}_{ij}$ and whose structural properties resemble those of the initial network \cite{Sonnenschein2012}. In particular, we assume that the weights of this new adjacency matrix are given by the probabilities of the corresponding nodes being connected, that is, \mbox{$\tilde{A}_{ij} = p_{ij}$}, where $p_{ij}$ is the probability of node~$i$, with degree~$k_i$, being connected to node~$j$, with degree~$k_j$.


For uncorrelated networks of the configuration ensemble, i.e., random networks with a given degree sequence $\{k_i\}$ and with a structural cutoff at \mbox{$k_i < \sqrt{N \overline{k}}$}, we can approximate the probability of two nodes~\mbox{$i,j$} being connected \cite{Newman2003B,Boguna2004,Sood2008,Bianconi2009} by
\begin{equation}
    p_{ij} \approx \frac{k_i k_j}{N \overline{k}} \, .
\end{equation}
In this way, we can approximate the sums over the neighbors of a given node $i$ as sums over the whole network,
\begin{equation}
    \sum_{j \in nn(i)} f_j = \sum_{j = 1}^N A_{ij} f_j \approx \sum_{j = 1}^N \frac{k_i k_j}{N \overline{k}} f_j \, ,
    \label{e:network_approximation}
\end{equation}
where $f_j$ is a function which can depend on the characteristics of node~$j$ ($k_j$ and/or $s_j$). Note that this approximation preserves the initial degree sequence, as it is obvious from
\begin{equation}
    k_i = \sum_{j=1}^N \tilde{A}_{ij} = k_i \frac{1}{\overline{k}} \left( \frac{1}{N} \sum_{j=1}^N k_j \right) \, ,
\end{equation}
and, therefore, the total number of links is also conserved.


\section{Results}


\subsection{Noise-induced, finite-size transition}


As shown in the previous literature about the Kirman model \cite{Kirman1993,Alfarano2008}, in the fully-connected case, the system is characterized by the existence of a finite-size transition between a bimodal and a unimodal behavior, depending on the relative magnitude of the noise and the interaction parameters. For $a < h/N$ the steady state probability distribution of $n$ is found to be bimodal with maxima at the extremes or fully ordered configurations, $n=0$ and $n=N$, meaning that, at any point in time, the most likely outcome of a static observation is to find a large majority of nodes in the same state, whether $0$ or $1$, with different observations leading to different predominant options (see \cite{carro2015} for an explanation in terms of an effective potential). On the contrary, for $a > h/N$ the distribution of $n$ becomes unimodal with a peak at $n = N/2$, meaning that, at any point in time, the most likely outcome of an observation is to find the system equally split between both options. Given the ergodicity of the model for $a \neq 0$, these probability distributions can also be understood in terms of the fractional time spent by the system with each value of $n$. In this manner, in the bimodal regime, stochastic realizations of the process will tend to be temporarily absorbed in the proximity of the fully ordered configurations with random switches between them, while realizations in the unimodal regime will spend most of the time with the system more or less equally divided among the two possible individual states, $0$ and $1$. At the critical point marking the transition between these two behaviors, $a_c = h/N$, the distribution of $n$ becomes uniform, meaning that any share of nodes between the two options is equally likely. Note that this transition is a finite-size effect, since the value of the critical point decreases for increasing system size and vanishes in the thermodynamic limit ($N \to \infty$).

\begin{figure}[ht!]
    \centering
    \includegraphics[width=11.5cm, height=!]{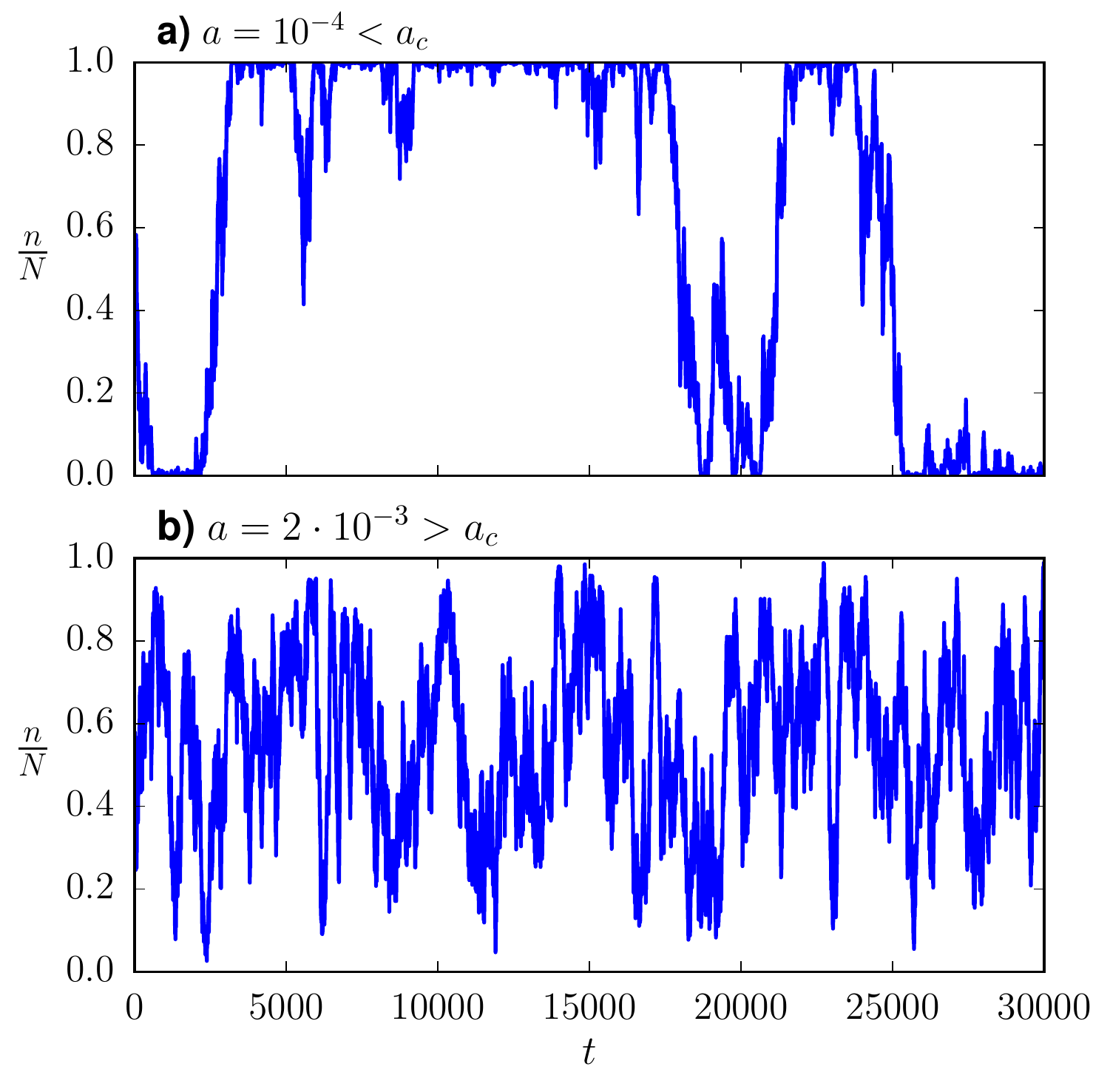}
    \caption{\label{f:magnetizationSingleRun} Fraction of nodes in state $1$ on a Barab\'asi-Albert scale-free network. Single realizations. The interaction parameter is fixed as $h = 1$, the system size as $N = 2500$ and the mean degree as $\overline{k} = 8$.}
    \end{figure}

The existence of the referred transition when the system is embedded in a network topology has also been reported in the literature both for the Kirman model \cite{Alfarano2009,Alfarano2011} and in the context of the noisy voter model \cite{Diakonova2015}. The above described phenomenology can thus also be observed in different network topologies. As an example, we show in Fig.~\ref{f:magnetizationSingleRun} two realizations of the dynamics for a Barab\'asi-Albert scale-free network corresponding, respectively, to the bimodal [panel \textbf{a)}] and the unimodal regime [panel \textbf{b)}]. A mean-field approach has been proposed in the literature \cite{Alfarano2009,Diakonova2015}, leading to an analytical solution for the critical point which does not depend on any property of the network other than its size, $a_c = h/N$, the transition still being a finite-size effect.

Both the analytical and numerical results to be presented here suggest, on the contrary, that the critical point does depend on the network, while they confirm the finite-size character of the transition. As a quantitative description of the transition we are going to use the variance of $n$: bearing in mind that the variance of a discrete uniform distribution between $0$ and $N$ is $N(N+2)/12$, we can identify the critical point of the transition as the relationship between the model parameters which leads the steady state variance of $n$ to take the value \mbox{$\sigma_{st}^2[n] = N(N+2)/12$}. Although it is not necessarily the case, numerical results confirm that the distributions obtained in this manner are indeed uniform.



\subsubsection*{Variance of $n$}


Introducing the annealed approximation for uncorrelated networks into the equation for the second-order cross-moments of the individual variables $s_i$, equation~\eqref{e:kirman_second_moment}, we can replace the sums over sets of neighbors by sums over the whole system. If we then rewrite this equation in terms of the covariance matrix $\sigma_{ij}$, defined as
\begin{equation}
    \sigma_{ij} = \langle s_i s_j \rangle - \langle s_i \rangle \langle s_j \rangle \, ,
\label{e:covariance_matrix_definition}
\end{equation}
we can use the relation
\begin{equation}
    \sigma^2 [n] = \langle n^2 \rangle - \langle n \rangle^2 = \sum_{ij} \langle s_i s_j \rangle - \sum_i \langle s_i \rangle \sum_j \langle s_j \rangle = \sum_{ij} \sigma_{ij}
    \label{e:variance_definition}
\end{equation}
to find an equation for the variance of $n$, by simply summing over $i$ and $j$. Finally, after some algebra (see Appendix~\ref{a:variance_of_n} for details), we find, in the steady state,
\begin{equation}
    \sigma^2_{st}[n] = \frac{N}{4} \left[ 1 + \frac{2h\left( 1 - \frac{1}{N} \right)}{4a + h} + \left( N - 3 + \frac{2}{N} \right) \frac{\displaystyle\left(\frac{h^2}{\overline{k}}\right) \overline{\left( \frac{k^2}{(4a+h)N\overline{k} + 2hk} \right)}}{2a + \displaystyle\left(\frac{h^2}{\overline{k}}\right) \overline{\left(\frac{k^2}{(4a+h) N \overline{k} + 2hk}\right)}} \right] \, ,
    \label{e:variance}
\end{equation}
under the necessary and sufficient condition that
\begin{equation}
    \forall i : k_i < \frac{(4a+h)N\overline{k}}{2h} \, ,
    \label{e:variance_condition}
\end{equation}
which is generally true and always true for $h>0$ and $\overline{k} \geq 2$. Note that the only approximation used in the derivation of equation~\eqref{e:variance} is the estimation of the adjacency matrix involved in the annealed approximation for uncorrelated networks.

\begin{figure}[ht!]
    \centering
    \includegraphics[width=12.5cm, height=!]{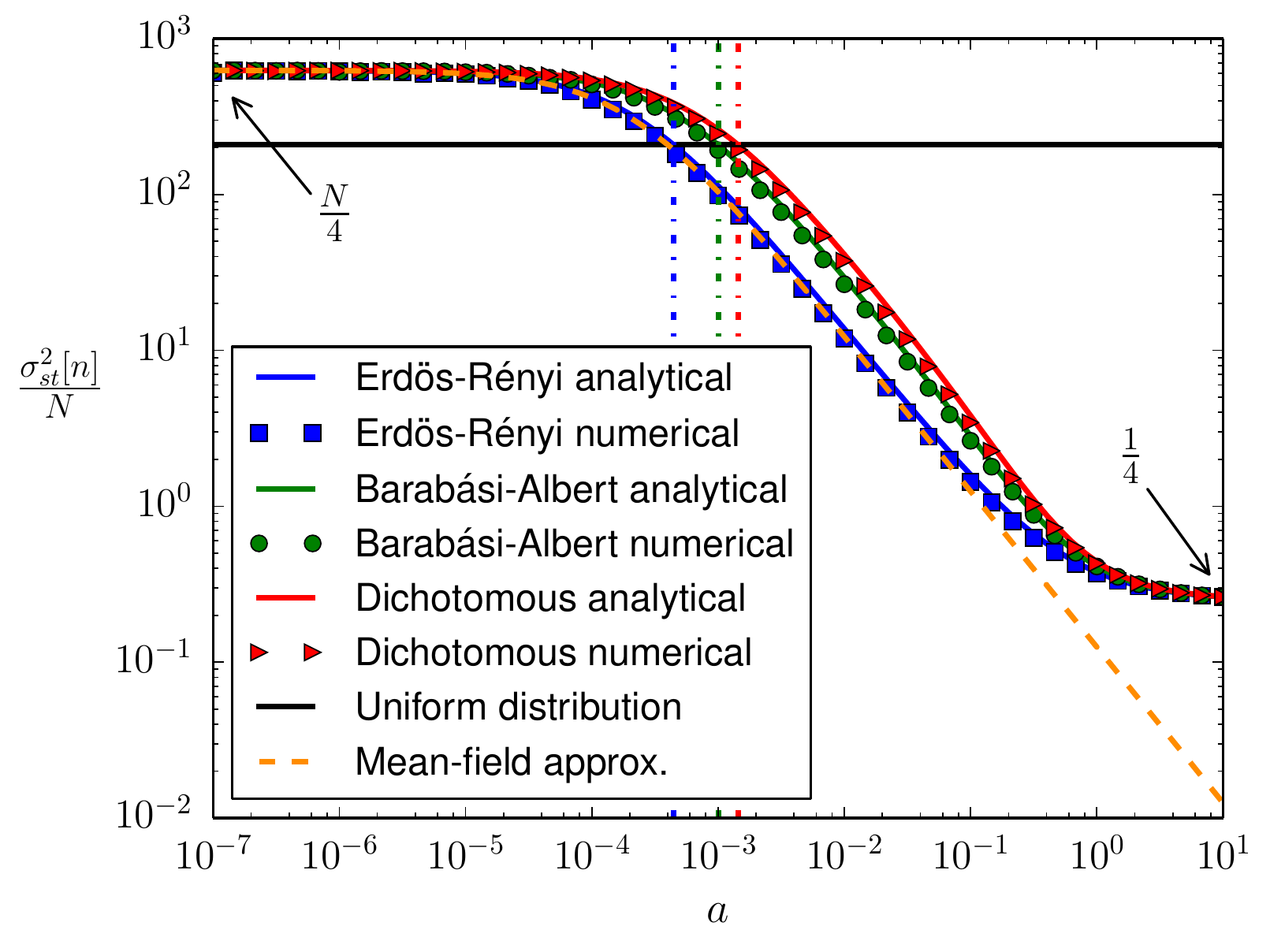}
    \caption{\label{f:varianceExact} Steady state variance of $n$ as a function of the noise parameter $a$, for three different types of networks: Erd\"os-R\'enyi random network, Barab\'asi-Albert scale-free network and dichotomous network. Symbols: Numerical results (averages over $20$ networks, $10$ realizations per network and $50000$ time steps per realization). Solid lines: Analytical results [see equation~\ref{e:variance}]. Dash-dotted lines: Analytical results for the critical points [see equation~\ref{e:critical_point}]. Dashed line: Mean-field approximation (see \cite{Alfarano2009}). The interaction parameter is fixed as $h = 1$, the system size as $N = 2500$ and the mean degree as $\overline{k} = 8$.}
    \end{figure}

The behavior of the variance $\sigma^2_{st}[n]$ as a function of the noise parameter $a$ is shown in Fig.~\ref{f:varianceExact} for the three types of networks studied. As we can observe, despite a small but systematic overestimation for intermediate values of the noise parameter ---attributable only to the annealed approximation for uncorrelated networks, the only one involved in its derivation---, the main features of the numerical steady state variance are correctly captured by the analytical expression in equation~\eqref{e:variance}. In particular, both its dependence on $a$ and the impact of the underlying network structure are well described by our approach. On the contrary, the mean-field solution proposed in the previous literature \cite{Alfarano2009}, and included in Fig.~\ref{f:varianceExact} for comparison, fails to reproduce the behavior of the variance of $n$ for large $a$ and is, by definition, unable to explain its dependence on the network topology. It is, nonetheless, a good approximation for the Erd\"os-R\'enyi random network and for values of the noise parameter~\mbox{$a \lesssim 10^{-1}$}.

Regarding the limiting behavior of the system when $a \to 0$ and when $a \to \infty$, we can observe, for both the numerical and the analytical results presented in Fig.~\ref{f:varianceExact}, that the influence of the network on the steady state variance of $n$ vanishes in both limits, where we recover the expected behaviors. Notably, in the limit of $a \to 0$ the variance tends to $N^2/4$ for all networks, and we progressively recover the voter model behavior; while in the limit of $a \to \infty$ the variance tends to $N/4$ regardless of the topology, as it corresponds to a purely noisy system composed by $N$ independent units adopting, randomly, values $0$ or $1$ (equivalent, as well, to a one dimensional random walk confined to the segment $[0,N]$).


Concerning the impact of the network structure, we can observe in Fig.~\ref{f:varianceExact} that for any finite value of the noise parameter, $0 < a < \infty$, a larger degree heterogeneity of the underlying topology, measured as the variance of the corresponding degree distribution, leads to a larger steady state variance of $n$. This behavior is further confirmed by the results to be presented in the next subsection, where we show the steady state variance of $n$ as a function of the variance of the underlying degree distribution $\sigma^2_k$, respectively, for two different values of the noise parameter $a$. As we can observe, even if the numerical results are systematically overestimated, our analytical approach [equation~\eqref{e:variance}] is able to capture the general features of this dependence and represents a significant improvement from the mean-field prediction of no network impact.


In order to study the bimodal-unimodal transition by using the behavior of the variance of $n$ illustrated in Fig.~\ref{f:varianceExact}, the variance value corresponding to a uniform distribution is included as a horizontal line, so that the critical $a$ value for each network can be easily identified at the corresponding intersection (marked by vertical dashed lines). Note that values of the variance of $n$ above (below) the uniform distribution line correspond to the system being in the bimodal (unimodal) phase. A first observation is that the referred transition still occurs when the noisy voter model is embedded in a network topology, thus confirming the results reported in the previous literature \cite{Alfarano2009,Diakonova2015}. However, as opposed to these previous studies, we can observe in Fig.~\ref{f:varianceExact} a clear dependence of the critical point on the underlying topology, an effect which seems to be correctly captured by our approach while it goes completely unnoticed, by definition, from a mean-field perspective. The particular features of this dependence will become clear by means of a first-order approximation of the steady state variance $\sigma^2_{st}[n]$ with respect to the system size $N$, allowing us to characterize the asymptotic behavior of the system for both small and large~$a$ as well as to find an explicit expression for the critical point $a_c$.


\subsubsection*{Asymptotic behavior of the variance of $n$}


Given that equation~\eqref{e:variance} does not allow for an intuitive analytical understanding of the network influence on the steady state variance of $n$, nor does it allow for an explicit analytical solution for the critical point $a_c$, we develop here a first-order approximation with respect to the system size $N$, which will also give a relevant insight regarding the asymptotic behavior of the system for both small and large~$a$. In fact, the result of this approximation strongly depends on the relationship between the system size $N$ and the noise parameter $a$, and we are thus led to consider two different approximation regimes.


In particular, when the noise parameter $a$ is of order $\mathcal{O}(N^{-1})$ or smaller, then the product $aN$ is, at most, of order $\mathcal{O}(N^0)$, and a first-order approximation of equation~\eqref{e:variance} with respect to the system size $N$ leads to
\begin{equation}
    \sigma^2_{st}[n] = \frac{N^2}{4} \left[ \frac{\displaystyle h \left( \frac{\sigma^2_k}{\overline{k}^2} + 1 \right)}{\displaystyle 2aN + h \left( \frac{\sigma^2_k}{\overline{k}^2} + 1 \right)} \right] + \mathcal{O}(N^{3/2}) \, ,
    \label{e:variance_small_a}
\end{equation}
corresponding to the asymptotic behavior of the variance of $n$ for small~$a$ and large~$N$. On the contrary, when $a$ is of order $\mathcal{O}(N^0)$ or larger, the product $aN$ is, at least, of order $\mathcal{O}(N)$, and the first-order approximation of equation~\eqref{e:variance} becomes
\begin{equation}
    \sigma^2_{st}[n] = \frac{N}{4} \left[ 1 + \frac{h}{2a} + \frac{\displaystyle h^2 \frac{\sigma^2_k}{\overline{k}^2} }{2a ( 4a + h)} \right] + \mathcal{O}(N^{1/2}) \, ,
    \label{e:variance_large_a}
\end{equation}
corresponding to the asymptotic behavior of the variance of $n$ for large~$a$ and large~$N$ (see Appendix~\ref{a:asymptotic_approximations_for_the_variance_of_n} for details).



\begin{figure}[ht!]
    \centering
    \includegraphics[width=12.5cm, height=!]{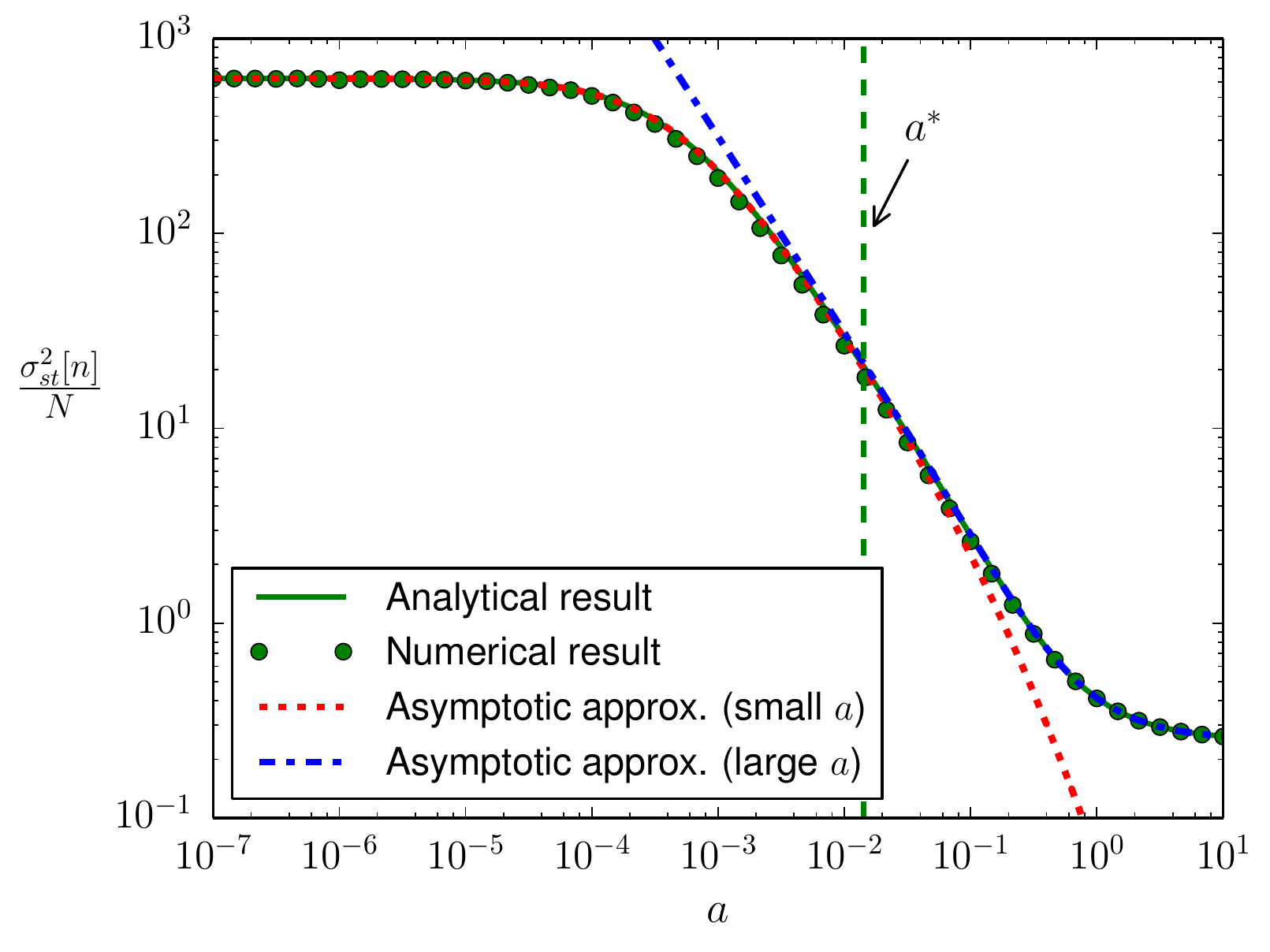}
    \caption{\label{f:varianceExactValidity} Steady state variance of $n$ as a function of the noise parameter $a$ for a Barab\'asi-Albert scale-free network. Symbols: Numerical results (averages over $20$ networks, $10$ realizations per network and $50000$ time steps per realization). Solid line: Analytical results [see equation~\eqref{e:variance}]. Dotted line: asymptotic approximation for small~$a$ [see equation~\eqref{e:variance_small_a}]. Dash-dotted line: asymptotic approximation for large~$a$ [see equation~\eqref{e:variance_large_a}]. Dashed line: Crossover point between both asymptotic approximations ($a^* = 0.014157$). The interaction parameter is fixed as $h = 1$, the system size as $N = 2500$ and the mean degree as $\overline{k} = 8$.}
    \end{figure}

For a more precise characterization of the ranges of validity of these two asymptotic approximations with respect to the noise parameter $a$, we present in Fig.~\ref{f:varianceExactValidity} the variance of $n$ as a function of $a$ for the numerical results and the three corresponding analytical expressions presented so far: the analytical result in equation~\eqref{e:variance}, the asymptotic expression for small~$a$ in equation~\eqref{e:variance_small_a} and the asymptotic expression for large~$a$ in equation~\eqref{e:variance_large_a}. Note the use a Barab\'asi-Albert scale-free network as an example. Furthermore, we also show in this figure the crossover point $a^*$ between both approximations, that we define as the value of $a$ that minimizes the distance between the logarithmic values of both functions~\eqref{e:variance_large_a} and~\eqref{e:variance_small_a}.

\begin{figure}[ht!]
    \centering
    \includegraphics[width=11.5cm, height=!]{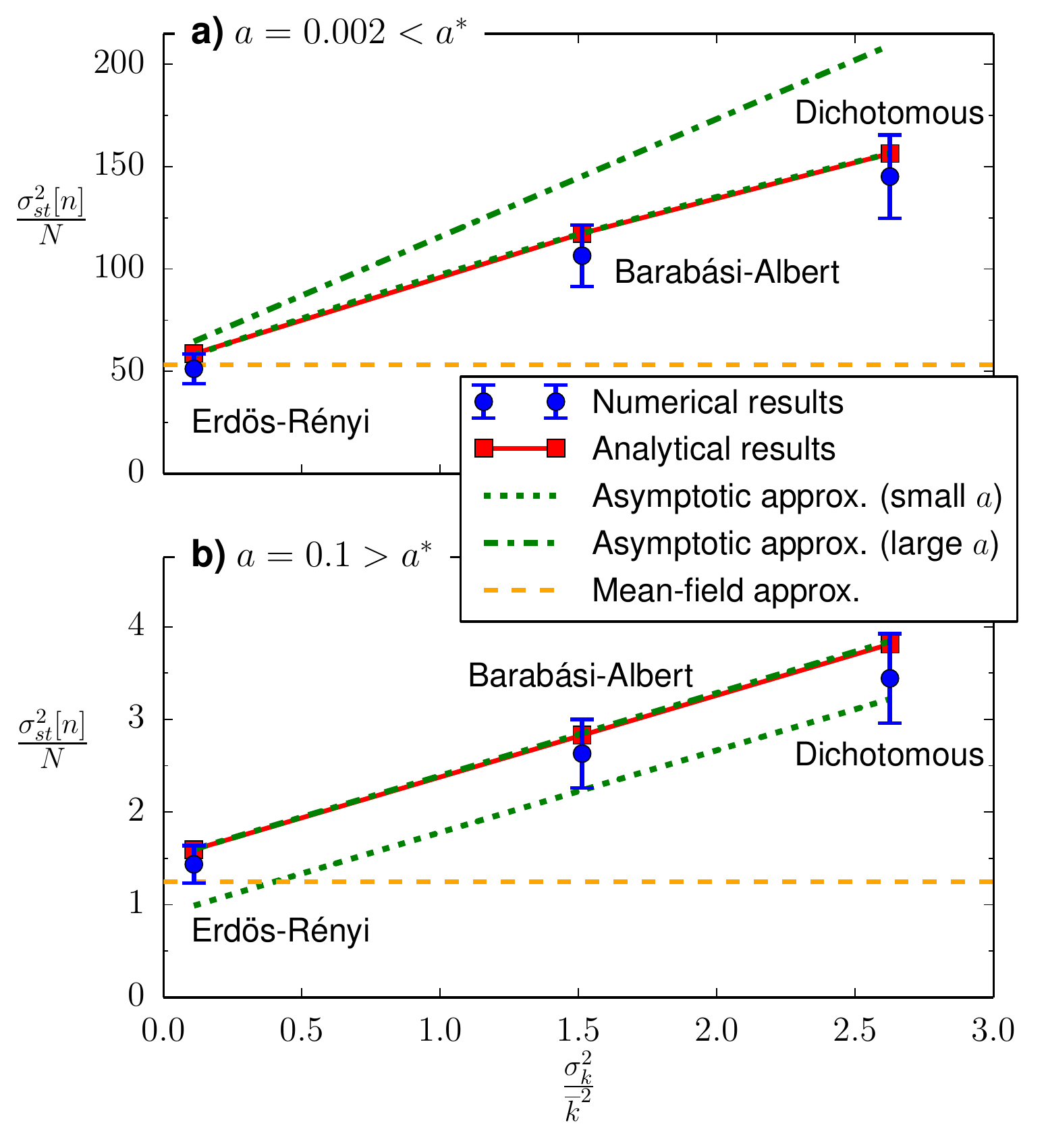}
    \caption{\label{f:varianceExactApprox2a} Steady state variance of $n$ as a function of the variance of the degree distribution $\sigma^2_k$ for two values of the noise parameter $a$. In order to keep all parameters constant except the variance of the degree distribution, a different network type is used for each point (in order of increasing $\sigma^2_k$: Erd\"os-R\'enyi random network, Barab\'asi-Albert scale-free network and dichotomous network). Circles with error bars: Numerical results (averages over $20$ networks, $10$ realizations per network and $50000$ time steps per realization). Solid line and squares: Analytical results [see equation~\eqref{e:variance}]. Dotted line: asymptotic approximation for small~$a$ [see equation~\eqref{e:variance_small_a}]. Dash-dotted line: asymptotic approximation for large~$a$ [see equation~\eqref{e:variance_large_a}]. Dashed line: Mean-field approximation (see \cite{Alfarano2009}). The interaction parameter is fixed as $h = 1$, the system size as $N = 2500$ and the mean degree as $\overline{k} = 8$.}
    \end{figure}

Noticing that, for both asymptotic approximations, the variance $\sigma^2_{st}[n]$ becomes an explicit function of the variance of the underlying degree distribution $\sigma^2_k$, we present in Fig.~\ref{f:varianceExactApprox2a} a comparison between these analytical functional relationships and the corresponding numerical results for two different values of the noise parameter $a$. In particular, taking into account the ranges of validity of the asymptotic approximations characterized above (see Fig.~\ref{f:varianceExactValidity}), we chose values of the noise parameter respectively before [panel~\textbf{a)}] and after [panel~\textbf{b)}] the crossover point $a^*$, and both of them in the region of $a$ leading to significant differences between network types (see Fig.~\ref{f:varianceExact}).

As we can observe, each asymptotic approximation accurately fits the analytical result in equation~\eqref{e:variance} within its respective range of validity, while it becomes clearly inaccurate out of this range. Therefore, we can use these approximations instead of equation~\eqref{e:variance} to better understand the behavior of the system. In this way, we can conclude that, regarding its impact on the results of the model, the most relevant property of the underlying network is not its mean degree, but the variance of its degree distribution relative to the square of its mean degree, $\sigma^2_k/\overline{k}^2$, a normalized measure of its degree heterogeneity. The results presented in Fig.~\ref{f:varianceExactApprox2a} show that this analysis significantly outperforms the mean-field prediction of no network impact, particularly for networks with large levels of degree heterogeneity. Note, nonetheless, that both asymptotic approximations are subject to the same inaccuracies in reproducing the numerical results as the original analytical expression, i.e., the inaccuracies caused by the annealed approximation for uncorrelated networks: a systematic overestimation of the numerical results and an inability to explain the results for topologies with large structural correlations.



\subsubsection*{Critical point}

\begin{figure}[ht!]
    \centering
    \includegraphics[width=11.5cm, height=!]{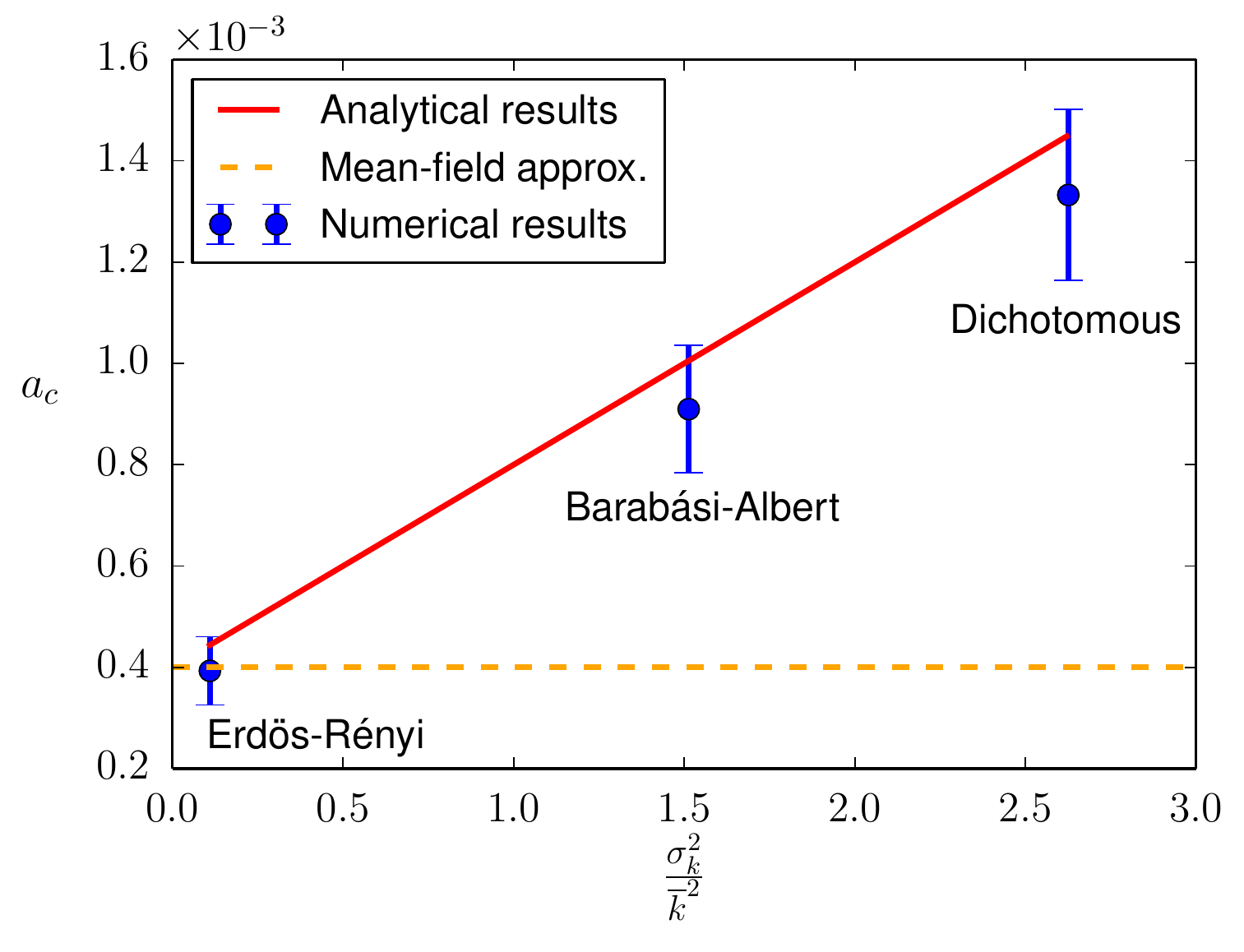}
    \caption{\label{f:criticalPoint} Critical value of the noise parameter $a$ as a function of the variance of the degree distribution of the underlying network, $\sigma_k^2$. In order to keep all parameters constant except the variance of the degree distribution, a different network type is used for each point (in order of increasing $\sigma^2_k$: Erd\"os-R\'enyi random network, Barab\'asi-Albert scale-free network and dichotomous network). Symbols: Numerical results (averages over $20$ networks, $10$ realizations per network and $50000$ time steps per realization). Solid line: Analytical results [see equation~\eqref{e:critical_point}]. Dashed line: Mean-field approximation (see \cite{Alfarano2009,Diakonova2015}). The interaction parameter is fixed as $h = 1$, the system size as $N = 2500$ and the mean degree as $\overline{k} = 8$.}
    \end{figure}

As described above, the critical point of the bimodal-unimodal transition can be defined as the relationship between the model parameters $a$ and $h$ leading the steady state variance of $n$ to take the value \mbox{$\sigma_{st}^2[n] = N(N+2)/12$}, which corresponds to a uniform distribution between $0$ and $N$. A numerical solution for the critical point $a_c$ can thus be found by applying this definition to the analytical expression for the variance of $n$ given in equation~\eqref{e:variance}. However, for a fully analytical description of the critical point, we have to use one of the asymptotic approximations presented above, algebraically solvable for $a_c$. In particular, bearing in mind that the value of the critical point of a fully-connected system is of order $\mathcal{O}(N^{-1})$ and that the change due to the network structure appears to be of order $\mathcal{O}(N^0)$ (see Fig.~\ref{f:varianceExact}), then we can expect the value of the critical point to be still of order $\mathcal{O}(N^{-1})$ and we can therefore use the small~$a$ asymptotic approximation in equation~\eqref{e:variance_small_a} to find
\begin{equation}
    a_c = \frac{h}{N} \left( \frac{\sigma^2_k}{\overline{k}^2} + 1 \right) + \mathcal{O}(N^{-3/2})\, ,
    \label{e:critical_point}
\end{equation}
to the first-order in $N$ (see Appendix~\ref{a:critical_point_approximation} for details). Both this expression and the mean-field approximation previously proposed in the literature \cite{Alfarano2009,Diakonova2015} are contrasted with numerical results in Fig.~\ref{f:criticalPoint}, where we present the values of the critical point $a_c$ for different types of networks as a function of the variance of the corresponding degree distributions $\sigma_k^2$.

As before, we notice in Fig.~\ref{f:criticalPoint} a systematic overestimation of the numerical results by our analytical approach, whose origin lies, again, in the annealed approximation for uncorrelated networks. While both equation~\eqref{e:critical_point} and the mean-field approximation are able to capture the finite-size character of the transition ---the fact that $a_c \to 0$ when $N \to \infty$---, only our approach is able to reproduce the influence of the underlying network structure on the critical point. In particular, we observe a numerical behavior approximately consistent with a linear relationship between the value of the critical point and the variance of the underlying degree distribution, as predicted by equation~\eqref{e:critical_point}. A quantitative assessment of the significance of this dependence can be obtained by observing the shift between the critical points corresponding to the Erd\"os-R\'enyi random network and the dichotomous network, the latter being almost a factor of $3$ larger than the former. While the persistence of the bimodal-unimodal, finite-size transition in different network topologies had already been reported \cite{Alfarano2009,Diakonova2015}, to the best of our knowledge, no dependence of the critical point on the characteristics of the underlying network has been documented so far for the noisy voter model nor in the context of the Kirman model (see \cite{Lambiotte2007} for a similar effect in a different model).


\subsection{Local order}


We can characterize the local order of the system with an order parameter $\rho$ defined as the interface density or density of active links, that is, the fraction of links connecting nodes in different states,
\begin{equation}
    \rho = \frac{\displaystyle\frac{1}{2} \sum_{i=1}^N A_{ij} [s_i(1-s_j) + (1-s_i)s_j]}{\displaystyle\frac{1}{2} \sum_{i=1}^N A_{ij}} \, ,
    \label{e:rho_definition}
\end{equation}
where $A_{ij}$ are the elements of the adjacency matrix. Larger values of $\rho$ imply a larger disorder, corresponding $\rho=1/2$ to a random distribution of states, while $\rho=0$ corresponds to full order. Furthermore, note that, as opposed to $n$, the order parameter does take into account the structure of connections between nodes.


While it has not been studied before in the context of the Kirman model, the interface density $\rho$ is commonly used to describe the time evolution of the voter model \cite{Suchecki2005B}. In the absence of noise, the voter model is characterized by the existence of two absorbing states ($n=0$ and $n=N$), both of them corresponding to full order ($\rho=0$). Therefore, the focus is on how the system approaches these absorbing ordered states. In the presence of noise, on the contrary, the system has no absorbing states, i.e., it is always active. Thus, the focus is not anymore on how it reaches any final configuration, but rather on characterizing its behavior once the influence of the initial condition has vanished, that is, in the steady state. In the context of the noisy voter model, it has been recently shown that, after a short initial transient, the average interface density reaches a plateau at a certain value $\langle \rho \rangle_{st}$, with \mbox{$\langle \rho \rangle_{st} > 0$} for any non-zero value of the noise and \mbox{$\langle \rho \rangle_{st} = 1/2$} in the infinite noise limit \cite{Diakonova2015}. Moreover, a mean-field pair-approximation has been used to find an analytical solution for $\langle \rho \rangle_{st}$ as a function of the level of noise and the mean degree of the underlying network. This analytical solution has been shown to be a good approximation for large values of the noise parameter, the small noise region not having been considered.



\begin{figure}[ht!]
    \centering
    \includegraphics[width=11.5cm, height=!]{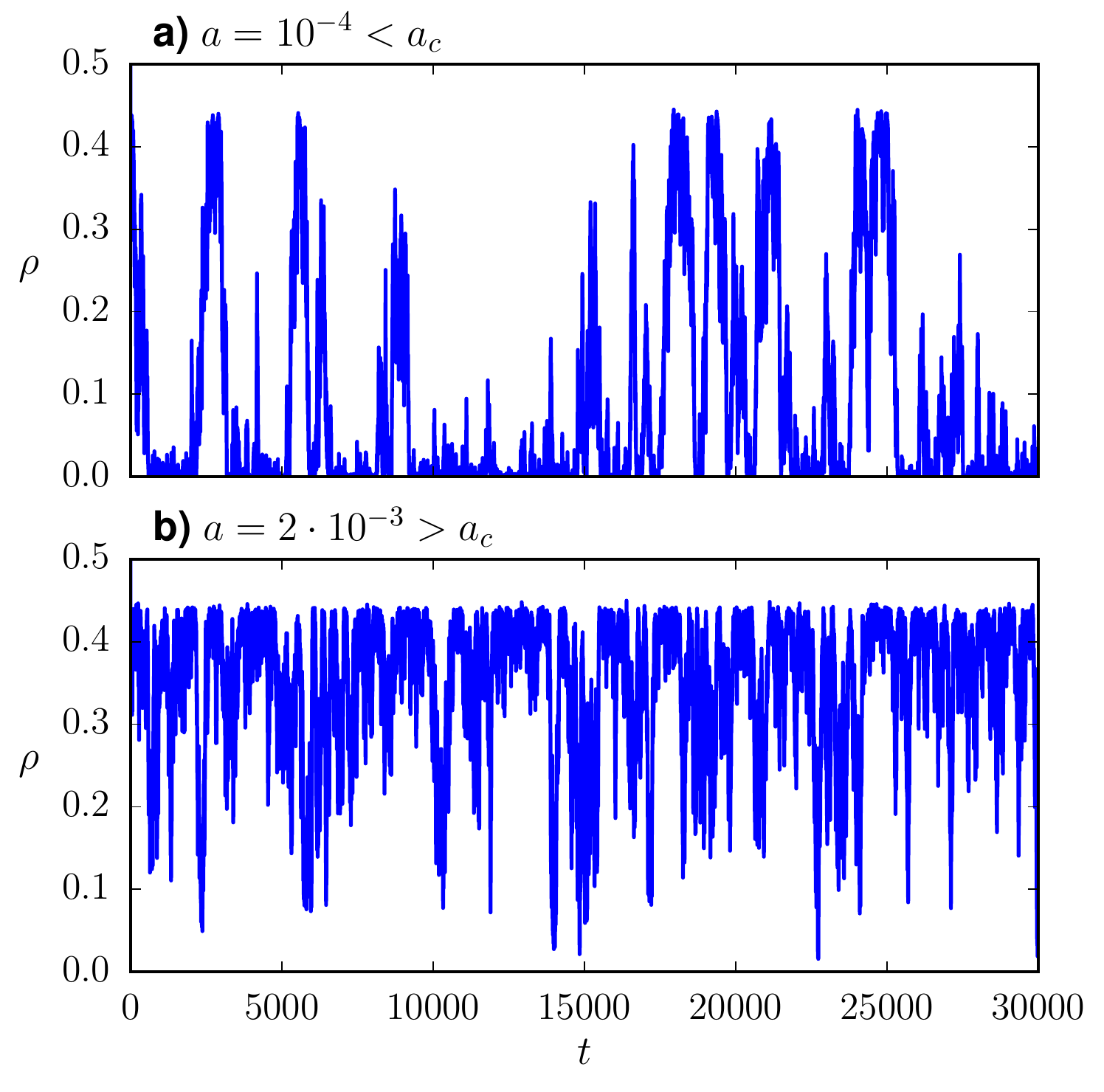}
    \caption{\label{f:initialOrderingSingleRun} Interface density on a Barab\'asi-Albert scale-free network. Single realizations (the same realizations shown in Fig.~\ref{f:magnetizationSingleRun}). The interaction parameter is fixed as $h = 1$, the system size as $N = 2500$ and the mean degree as $\overline{k} = 8$.}
    \end{figure}

Let us start the description of our results by emphasizing that individual realizations of the interface density $\rho$ remain always active for any non-zero value of the noise, as it was also the case for the variable $n$ (see Fig.~\ref{f:magnetizationSingleRun}). As an example, we show in Fig.~\ref{f:initialOrderingSingleRun} two realizations of the dynamics for a Barab\'asi-Albert scale-free network corresponding, respectively, to the bimodal [panel \textbf{a)}] and the unimodal regime [panel \textbf{b)}]. While in the first of them ($a < a_c$) the system fluctuates near full order, with sporadic excursions of different duration and amplitude towards disorder; in the second ($a > a_c$), the system fluctuates around a high level of disorder, with some large excursions towards full order.

\begin{figure}[ht!]
    \centering
    \includegraphics[width=14.5cm, height=!]{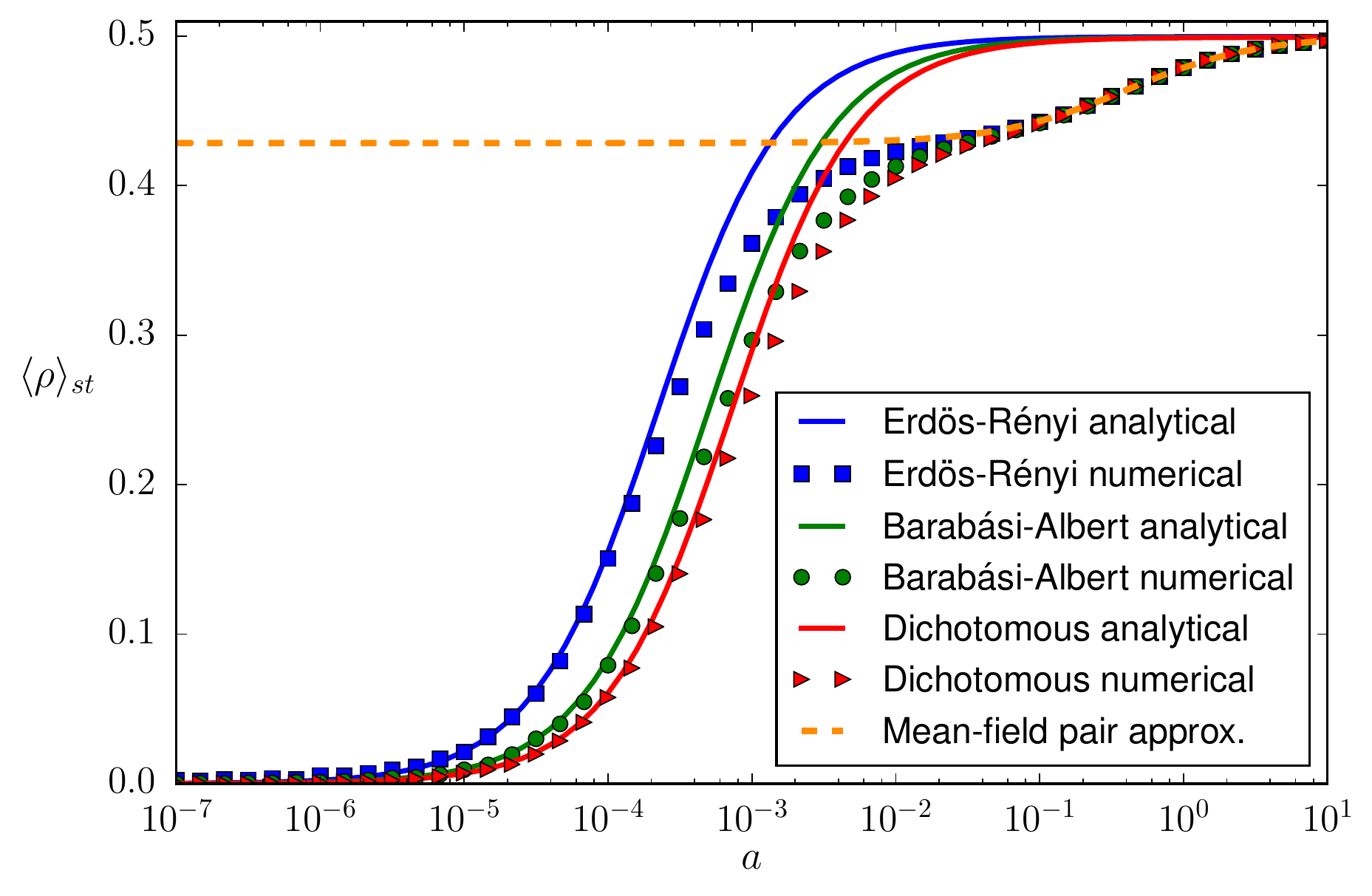}
    \caption{\label{f:rhoExactLogX} Steady state of the average interface density as a function of the noise parameter $a$ in a linear-logarithmic scale and for three different types of networks: Erd\"os-R\'enyi random network, Barab\'asi-Albert scale-free network and dichotomous network. Symbols: Numerical results (averages over $20$ networks, $10$ realizations per network and $50000$ time steps per realization). Solid lines: Analytical results [see equation~\eqref{e:rho}]. Dashed line: Mean-field pair-approximation (see \cite{Diakonova2015}). The interaction parameter is fixed as $h = 1$, the system size as $N = 2500$ and the mean degree as $\overline{k} = 8$.}
    \end{figure}

Introducing the annealed approximation for uncorrelated networks described above into the definition of the order parameter given in equation~\eqref{e:rho_definition}, and focusing on the steady state average value, we obtain
\begin{equation}
    \langle \rho \rangle_{st} = \sum_{ij} \frac{k_i k_j}{(N \overline{k})^2} \Big( \langle s_i \rangle_{st} + \langle s_j \rangle_{st} - 2 \langle s_j s_j \rangle_{st} \Big) \, .
    \label{e:rho_0}
\end{equation}
In this way, an explicit solution for the steady state average interface density can be found by expressing it in terms of the analytical results presented so far, namely, in terms of the variance $\sigma^2_{st}[n]$ (see Appendix~\ref{a:order_parameter_the_interface_density} for details),
\begin{equation}
    \langle \rho \rangle_{st} = \frac{1}{2} - \frac{2}{(hN)^2} \left[ \frac{(4a+h)(2a+h)}{\left( 1 - \frac{1}{N} \right) \left( 1 - \frac{2}{N} \right)} \left( \sigma^2[n] - \frac{N}{4} \right) - \frac{\left( a + \frac{h}{2} \right)}{\left( 1 - \frac{2}{N} \right)} hN \right] \, .
    \label{e:rho}
\end{equation}
This expression can be contrasted with numerical results in Fig.~\ref{f:rhoExactLogX}, where we present the steady state average interface density $\langle \rho \rangle_{st}$ as a function of the noise parameter $a$ for different types of networks. The mean-field pair-approximation result derived in \cite{Diakonova2015} is also included for comparison.

As we can observe in Fig.~\ref{f:rhoExactLogX}, our approach correctly captures the behavior of the system for both small ($a \lesssim 10^{-3}$) and very large values ($a \gtrsim 3$) of the noise parameter: both the asymptotic convergence towards \mbox{$\langle \rho \rangle_{st} = 0$} for small $a$ (voter model result for finite systems) and the convergence towards \mbox{$\langle \rho \rangle_{st} = 1/2$} for large $a$ (full disorder) are well reproduced. On the contrary, our analytical approach fails to reproduce the numerical results for intermediate values of the noise parameter (\mbox{$10^{-3} \lesssim a \lesssim 3$}). The origin of this discrepancy lies in the annealed network approximation: when replacing the original network by a weighted fully-connected topology, all track of local effects is lost ---precisely those measured by the order parameter. The fact that this discrepancy is only present for intermediate values of $a$ can be explained, on the one hand, by the lack of locally ordered structures in the fully disordered, large $a$ regime and, on the other hand, by the development of a global order ---more and more independent of local effects--- for decreasing values of $a$. Thus, an accurate fit of the numerical results for any value of $a$ can only be expected for topologies where local effects are absent or negligible. In the Supplementary Figure S1 presented in Appendix~\ref{a:suppementary_Figure_S1} we show that the approximation successfully fits the results in a fully-connected network. The good accuracy of the results presented above for the variance of $n$ suggests that the discrepancy between analytical and numerical results appears only when the annealed network approximation is used to derive a relationship between $\langle \rho \rangle_{st}$ and $\sigma^2_{st}[n]$, and not in the derivation of the latter, for which only global correlations are relevant. Apart from the functional dependence of the interface density on the noise parameter, our approach is also able to capture the influence of the network, which becomes significant for $a \lesssim 10^{-2}$. In particular, we find that a larger variance of the degree distribution of the corresponding network leads to a smaller interface density, i. e., to a higher level of order.


Even if the mean-field pair-approximation fits the numerical results remarkably well for large and intermediate values of the noise parameter ($a \gtrsim 10^{-2}$), it is completely unable to reproduce the behavior of the system for small $a$, and it fails to explain the influence of any network property other than the mean degree. While the pair-approximation allows to capture the short-range order characteristic of intermediate values of $a$, the assumptions implicit in the derivation of the mean-field result \cite{Vazquez2008B} do not allow to reproduce the long-range order characteristic of the small $a$ region. Note that the limiting case of $a=0$ (voter model) is a singular point of the mean-field pair-approximation \cite{Diakonova2015}, leading to the existence of two different solutions: a non-zero solution linked to the result displayed in Fig.~\ref{f:rhoExactLogX} ---correct in the infinite size limit---, and a zero solution ---correct for finite systems.


\subsection{Inference of network properties from the autocorrelation of $n$}


As explained above, from any initial condition, the system quickly reaches a dynamic steady state, whose active character can be clearly observed in Fig.~\ref{f:magnetizationSingleRun}. In order to characterize the dynamic nature of this steady state, let us now focus on the steady state autocorrelation function of $n$, defined as
\begin{equation}
    K_{st}[n](\tau) = \langle n(t + \tau) n(t) \rangle_{st} - \langle n \rangle_{st}^2 \, ,
    \label{e:autocorrelation_definition}
\end{equation}
where $\tau$ plays the role of a time-lag. In the fully-connected case, it has been shown in the previous literature \cite{Alfarano2008} that the autocorrelation decays exponentially, with an exponent proportional to the noise parameter, \mbox{$K_{st}[n](\tau) = \sigma^2_{st}[n] e^{-2a\tau}$}. In the case of different network topologies, the mean-field prediction is that no influence of the network is to be expected and, therefore, the same exponential decay as in the fully-connected case is to be found. In contrast with this prediction, both the analytical and numerical results to be presented here show that the network does have a significant impact on the functional form of the steady state autocorrelation of $n$.



\begin{figure}[ht!]
    \centering
    \includegraphics[width=12.2cm, height=!]{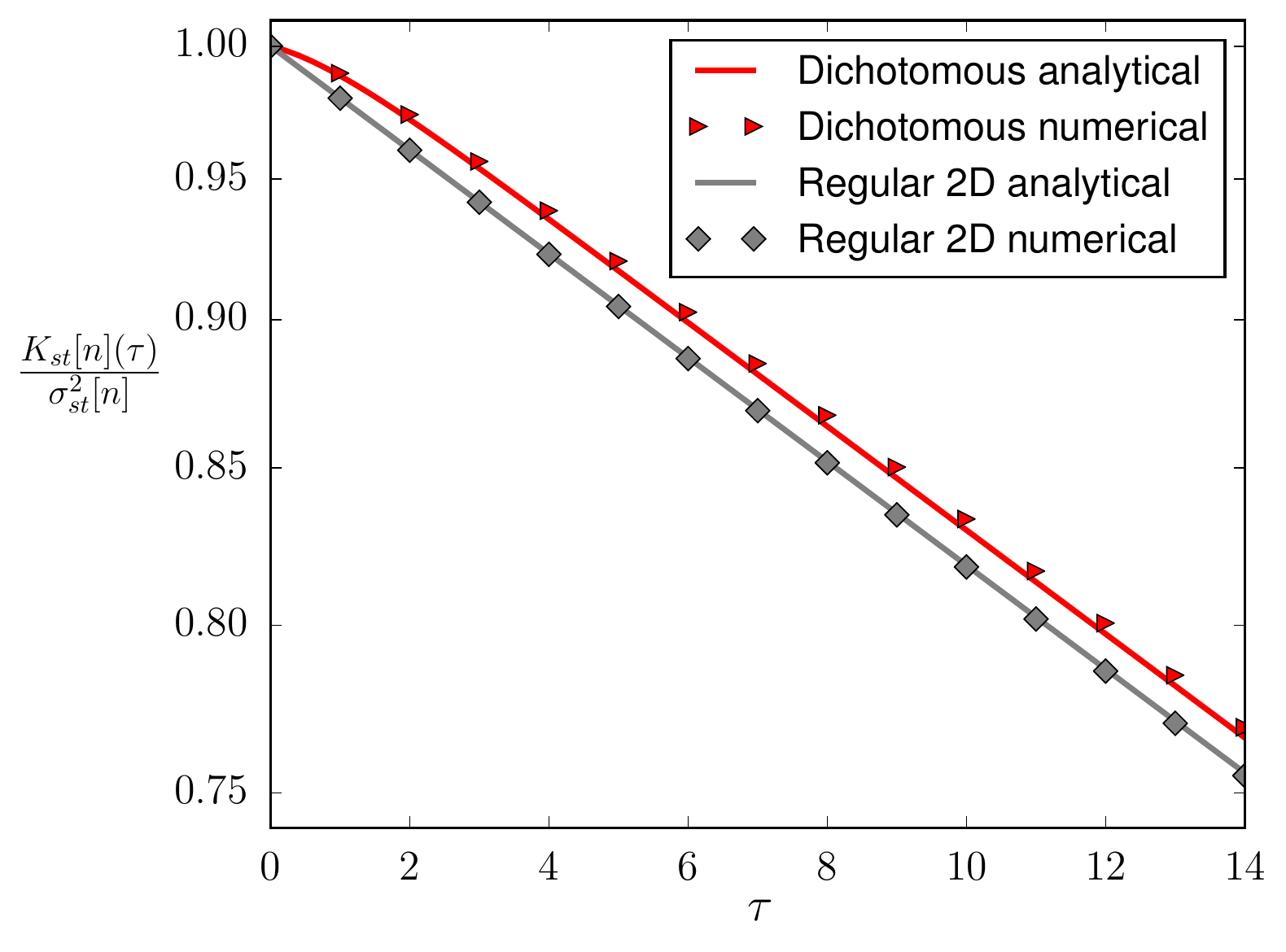}
    \caption{\label{f:autoCorrExact2Nets} Autocorrelation function of $n$ in log-linear scale for a dichotomous network and a regular 2D lattice. Symbols: Numerical results (averages over $10$ networks, $2$ realizations per network and $200000$ time steps per realization). Solid lines: Analytical results [see equation~\eqref{e:autocorrelation}]. Parameter values are fixed as $a = 0.01$, $h = 1$, the system size as $N = 2500$ and the mean degree as $\overline{k} = 8$.}
    \end{figure}

Introducing the annealed approximation for uncorrelated networks described above into the equation for the time evolution of the first-order moments~\eqref{e:kirman_first_moment}, integrating it with carefully chosen initial conditions, and making use of the above reported analytical results (see Appendix~\ref{a:autocorrelation_function_of_n} for details), we can find
\begin{equation}
    K_{st}[n](\tau) = \left( \sigma^2_{st}[n] - S_1 \right) e^{-(2a + h) \tau} + S_1 e^{-2a \tau} \, ,
    \label{e:autocorrelation}
\end{equation}
where $S_1$ is defined as
\begin{equation}
    S_1 = \frac{2a + h}{h \left( 1 - \frac{1}{N} \right)} \left( \sigma^2_{st}[n] - \frac{N}{4} \right) \, .
    \label{e:autocorrelation_C}
\end{equation}
This expression can be contrasted with numerical results in Fig.~\ref{f:autoCorrExact2Nets}, where we present the autocorrelation function, normalized by the variance, for the two extreme cases of a network with no degree heterogeneity (regular 2D lattice) and a highly heterogeneous degree distribution (dichotomous network). Note the logarithmic scale in the y-axis.

It is important to note that, in the case of no degree heterogeneity, the new variable $S_1$ becomes \mbox{$S_1 = \sigma^2_{st}[n]$}. This can be understood by applying, $\forall i$, $k_i = \overline{k}$ in the averages over the degree distribution in equation~\eqref{e:variance},
\begin{equation}
    \sigma^2_{st}[n] = \frac{N}{4} \frac{(2a + h)}{\left( 2a + \frac{h}{N} \right)} \, ,
\end{equation}
and introducing this result into the definition of $S_1$, equation~\eqref{e:autocorrelation_C}. Thus, for networks with no degree heterogeneity, the steady state autocorrelation function behaves as in the fully-connected case, and as predicted by the mean-field approximation, \mbox{$K_{st}[n](\tau) = \sigma^2_{st}[n] e^{-2a\tau}$}. This single exponential decay is confirmed by the numerical results presented in Fig.~\ref{f:autoCorrExact2Nets} for the regular 2D lattice.


On the contrary, for networks with non-zero degree heterogeneity, in general, \mbox{$S_1 \neq \sigma^2_{st}[n]$}, and thus the autocorrelation function consists of two different exponential decay components [see equation~\eqref{e:autocorrelation}]. When \mbox{$h > 0$}, the exponential $e^{-(2a+h)\tau}$ decays faster than $e^{-2a\tau}$. Therefore, for long time-lags, we expect the normalized autocorrelation function of any network to be parallel to $e^{-2a\tau}$ in log-linear scale, with a vertical shift proportional to its degree heterogeneity and due to the initial deviation from the single exponential behavior. This description is confirmed by the numerical results presented in Fig.~\ref{f:autoCorrExact2Nets} for the dichotomous network.



Neither equation~\eqref{e:variance} nor the asymptotic approximate expressions~\eqref{e:variance_small_a} and~\eqref{e:variance_large_a} allow to infer, for a given system, the values of the two model parameters, $a$ and $h$, and the normalized variance of the underlying degree distribution, $\sigma^2_k/\overline{k}^2$, by measuring only the steady state variance of $n$, $\sigma^2_{st}[n]$. Thus, it is impossible, by using only these relationships, to conclude if the fluctuations observed in a given system have a contribution due to the degree heterogeneity of the network, without a prior knowledge of the model parameters $a$ and $h$. On the contrary, the particular functional form of the autocorrelation function $K_{st}[n](\tau)$ ---with two exponential decay components whose exponents are different functions of $a$ and $h$--- does allow for the values of $a$, $h$ and $\sigma^2_k/\overline{k}^2$ to be inferred from equation~\eqref{e:autocorrelation}, in combination with equation~\eqref{e:variance_small_a} or equation~\eqref{e:variance_large_a}, by measuring only the temporal correlations of the aggregated variable $n$, and assuming we can also know the system size $N$. Note that the use of equation~\eqref{e:variance_small_a} or equation~\eqref{e:variance_large_a} can be determined by self-consistency, depending on the value obtained for $a$. As an example, a simple fit of equation~\eqref{e:autocorrelation} to the numerical results presented in Fig.~\ref{f:autoCorrExact2Nets} for the dichotomous network leads, in combination with equation~\eqref{e:variance_small_a}, to the fitted parameter values $a = 0.0099$, $h = 0.94$ and \mbox{$\sigma^2_k/\overline{k}^2 = 2.539$}, remarkably close to the actual values used for computing the numerical results, $a = 0.01$, $h = 1$ and \mbox{$\sigma^2_k/\overline{k}^2 = 2.625$}. In this way, we are able to infer some information about the underlying network ---its normalized level of degree heterogeneity--- by studying only the aggregate behavior of the system as a whole.


\section*{Discussion}



In this paper, we have proposed a new analytical method to study stochastic, binary-state models of interacting units on complex networks. Moving beyond the usual mean-field theories \cite{Vazquez2008B,Alfarano2009,Diakonova2015}, this alternative approach builds on a recent study considering heterogeneity in stochastic interacting particle systems \cite{Lafuerza2013} and proposes an annealed approximation for uncorrelated networks accounting for the network structure as parametric heterogeneity.


Using the noisy voter model as an example, we have been able to unfold the dependence of the model not only on the mean degree of the underlying topology (the mean-field prediction) but also on more complex averages over the degree distribution. In particular, we have shown that the degree heterogeneity ---i.e., the variance of the underlying degree distribution--- has a substantial influence on the location of the critical point of the noise-induced, finite-size transition characterizing the model. This shift of the transition might have important practical implications in real systems, since it suggests that different behavioral regimes can be achieved by introducing changes in the underlying network of interactions. Furthermore, we have studied the influence of the network on the local ordering of the system, finding that a larger degree heterogeneity leads to a higher average level of order in the steady state. Interestingly, we have also found the heterogeneity of the underlying degree distribution to play a relevant role in determining the functional form of the temporal correlations of the system. Finally, we have shown how this latter effect can be used to infer some information about the underlying network ---its normalized level of degree heterogeneity--- by studying only the aggregate behavior of the system as a whole, an issue of interest for systems where macroscopic, population level variables are easier to measure than their microscopic, individual level counterparts.


Numerical simulations on different types of networks have been used to validate our analytical results, finding a remarkably good agreement for all the properties studied except for the local order, for which a significant discrepancy is found for intermediate levels of noise. The origin of this discrepancy has been shown to lie in the annealed network approximation, whose validity is restricted to global properties or situations where local effects are negligible. The generally good agreement found is all the more remarkable considering that, while the uncorrelated network assumption is essential for the proposed analytical method, we did not impose any particular structural constraint to avoid correlations in the networks used for the numerical simulations \cite{Boguna2004,Catanzaro2005}.


\begin{appendices}

\renewcommand{\thesubsection}{\Alph{subsection}}
\renewcommand{\thesection}{}
\titleformat{\section}{\Large\sffamily\bfseries}{}{0cm}{#1}{}
\titleformat{\subsection}{\large\sffamily\bfseries}{\thesubsection}{0.5em}{#1}{}
\titlespacing*{\subsection}{0pc}{3ex plus 4pt minus 3pt}{5pt}
\numberwithin{equation}{subsection}
\renewcommand{\theequation}{\Alph{subsection}\arabic{equation}}

\section{Appendices}

\subsection{Master equation}
\label{a:master_equation}

We derive here a general master equation for the \mbox{$N$-node} probability distribution \mbox{$P(s_1,\ldots,s_N)$}, where the individual node variables are binary and take the values \mbox{$s_i=\{0,1\}$}. Recalling that $r_i^+$ is the rate at which node $i$ changes its state from $s_i=0$ to $s_i=1$ and $r_i^-$ the rate at which it does so in the opposite direction, we can directly write differential equations for the probability of node $i$ to be in state $s_i=0$ and for its probability to be in state $s_i=1$, respectively,
\begin{equation}
\begin{aligned}
    \frac{dP(s_i=0)}{dt} = -r_i^+ P(s_i=0) + r_i^- P(s_i=1) \, ,\\[5pt]
    \frac{dP(s_i=1)}{dt} = -r_i^- P(s_i=1) + r_i^+ P(s_i=0) \, .
    \label{ae:master_equation_1}
\end{aligned}
\end{equation}
Introducing here the individual-node step operators \mbox{$E_i^{+1}$} and \mbox{$E_i^{-1}$}, whose effect over an arbitrary function of the state of node $i$, $f(s_i)$, is defined as
\begin{equation}
\begin{aligned}
    &E_i^{+1} \bigl[f(s_i = 0)\bigr] = f(s_i = 1) \, ,\\[3pt]
    &E_i^{+1} \bigl[f(s_i = 1)\bigr] = 0 \, ,\\[3pt]
    &E_i^{-1} \bigl[f(s_i = 0)\bigr] = 0 \, ,\\[3pt]
    &E_i^{-1} \bigl[f(s_i = 1)\bigr] = f(s_i = 0) \, ,
    \label{ae:step_operators}
\end{aligned}
\end{equation}
we can rewrite equations~\eqref{ae:master_equation_1} as
\begin{equation}
\begin{aligned}
    \frac{dP(s_i=0)}{dt} &= -r_i^+ P(s_i=0) + r_i^- E_i^{+1} P(s_i=0) \, ,\\[5pt]
    \frac{dP(s_i=1)}{dt} &= -r_i^- P(s_i=1) + r_i^+ E_i^{-1} P(s_i=1) \, .
    \label{ae:master_equation_2}
\end{aligned}
\end{equation}
Multiplying these two equations, respectively, by \mbox{$(1 - s_i)$} and $s_i$, we can gather them in a single differential equation,
\begin{equation}
    \frac{dP(s_i)}{dt} = (1 - s_i) \left[ -r_i^+ P(s_i) + r_i^- E_i^{+1} P(s_i) \right] + s_i \left[ -r_i^- P(s_i) + r_i^+ E_i^{-1} P(s_i) \right] \, ,
    \label{ae:master_equation_3}
\end{equation}
and noticing that \mbox{$(1 - s_i) = E_i^{+1} [s_i]$} and \mbox{$s_i = E_i^{+1} [(1 - s_i)]$}, we can rearrange terms as
\begin{equation}
    \frac{dP(s_i)}{dt} = \left( E_i^{+1} - 1 \right) \left[ s_i r_i^- P(s_i) \right] + \left( E_i^{-1} - 1 \right) \left[ (1 - s_i) r_i^+ P(s_i) \right] \, .
    \label{ae:master_equation_4}
\end{equation}
Finally, we find the master equation for the \mbox{$N$-node} probability distribution \mbox{$P(s_1,\ldots,s_N)$} by simply adding up the contribution of every single node \mbox{$i \in [1,N]$},
\begin{equation}
    \frac{dP(s_1,\ldots,s_N)}{dt} = \sum_{i=1}^N \left(E_i^{+1} - 1\right) \left[s_i r_i^- P(s_1,\ldots,s_N)\right] + \sum_{i=1}^N \left(E_i^{-1} - 1\right) \left[(1 - s_i) r_i^+ P(s_1,\ldots,s_N)\right] \, .
    \label{ae:master_equation}
\end{equation}


\subsection{Equation for the time evolution of the first-order moments $\langle s_i \rangle$}
\label{a:equation_for_the_time_evolution_of_the_first-order_moments}

We show, in this section, how to obtain a general equation for the time evolution of the first-order moments $\langle s_i \rangle$ [equation~\eqref{e:first_moment} in the main text]. Let us start by using the definition of the step operators in equation~\eqref{ae:step_operators} and the binary character of each individual node state variable, \mbox{$s_i = \{0,1\}$}, to derive, for a given function of the state of node $i$, \mbox{$f(s_i)$}, four relations which will ease later calculations. While the function $f$ might also depend on the other variables, \mbox{$f = f(s_1,...,s_i,...,s_N)$}, we restrict our attention, without loss of generality, to the case \mbox{$f(s_i)$}. For the first two relations, we have that
\begin{equation}
    \sum_{s_i} \left( E_i^{+1} - 1 \right) \left[ s_i f(s_i) \right] = \sum_{s_i} \left( E_i^{+1} \left[ s_i f(s_i) \right] - s_i f(s_i) \right) = 1 \cdot f(1) - 0 \cdot f(0) + 0 - 1 \cdot f(1) = 0 \, ,
\label{ae:relation_1}
\end{equation}
and
\begin{equation}
    \sum_{s_i} \left( E_i^{-1} - 1 \right) \left[ (1 - s_i) f(s_i) \right] = \sum_{s_i} \left( E_i^{+1} \left[ (1 - s_i) f(s_i) \right] - (1 - s_i) f(s_i) \right) = 0 - 1 \cdot f(0) + 1 \cdot f(0) - 0 \cdot f(1) = 0 \, ,
\label{ae:relation_2}
\end{equation}
where the sums are over the two possible values of $s_i$. Looking at the master equation~\eqref{ae:master_equation}, one can understand that these two relations translate the fact that any increase in the probability of a given node being in a given state must be accompanied by a corresponding decrease in the probability of the complementary state. Regarding the other two relations, we can write
\begin{equation}
\begin{aligned}
    \sum_{s_i} s_i \left( E_i^{+1} - 1 \right) \left[ s_i f(s_i) \right] &= \sum_{s_i} s_i \left( E_i^{+1} \left[ s_i f(s_i) \right] - s_i f(s_i) \right)\\[5pt]
    &= 0 \cdot (1 \cdot f(1) - 0 \cdot f(0)) + 1 \cdot (0 - 1 \cdot f(1)) = - 1 \cdot f(1)\\[5pt]
    &= - \sum_{s_i} s_i f(s_i) \, ,
\end{aligned}
\label{ae:relation_3}
\end{equation}
and
\begin{equation}
\begin{aligned}
    \sum_{s_i} s_i \left( E_i^{-1} - 1 \right) \left[ (1 - s_i) f(s_i) \right] &= \sum_{s_i} s_i \left( E_i^{+1} \left[ (1 - s_i) f(s_i) \right] - (1 - s_i) f(s_i) \right)\\[5pt]
    &= 0 \cdot (0 - 1 \cdot f(0)) + 1 \cdot (1 \cdot f(0) - 0 \cdot f(1)) = 1 \cdot f(0)\\[5pt]
    &= \sum_{s_i} (1 - s_i) f(s_i) \, .
\end{aligned}
\label{ae:relation_4}
\end{equation}

Let us also introduce, for clarity, the notation \mbox{$\sum_{\{s\}}$} to refer to the sum over all the possible combinations of states of all the individual nodes' variables,
\begin{equation}
    \sum_{\{s\}} \equiv \sum_{s_1} \sum_{s_2} \cdots \sum_{s_N} \, ,
    \label{ae:notation_sum}
\end{equation}
and \mbox{$\sum_{\{s\}_j}$} to indicate the sum over all the possible combinations of states of all the variables except $s_j$,
\begin{equation}
    \sum_{\{s\}_j} \equiv \sum_{s_1} \cdots \sum_{s_{j-1}} \sum_{s_{j+1}} \cdots \sum_{s_N} \, .
    \label{ae:notation_sum_excluding}
\end{equation}
Note that these two definitions are related by
\begin{equation}
    \sum_{\{s\}} = \sum_{\{s\}_j} \sum_{s_j} \, ,
    \label{ae:notations_relation}
\end{equation}
which allows us to split the sum over all possible configurations of the system into a sum over the values of one of the variables and a sum over the configurations of the rest of the system. By using the notation in~\eqref{ae:notation_sum}, the average of a given function of the states of the nodes, \mbox{$f(s_1,\ldots,s_N)$}, can be written as
\begin{equation}
    \Bigl\langle f(s_1,\ldots,s_N) \Bigr\rangle = \sum_{\{s\}} f(s_1,\ldots,s_N) P(s_1,\ldots,s_N) \, .
    \label{ae:average_definition}
\end{equation}

Using this expression and the master equation in~\eqref{ae:master_equation} we derive an equation for the time evolution of the average value of the state of node $i$,
\begin{equation}
\begin{aligned}
    \frac{d \langle s_i \rangle}{dt} &= \sum_{\{s\}} s_i \frac{dP(s_1,\ldots,s_N)}{dt}\\[5pt]
    &= \sum_{\{s\}} \sum_{j=1}^N s_i \left(E_j^{+1} - 1\right) \left[s_j r_j^- P(s_1,\ldots,s_N)\right] + \sum_{\{s\}} \sum_{j=1}^N s_i \left(E_j^{-1} - 1\right) \left[(1 - s_j) r_j^+ P(s_1,\ldots,s_N)\right] \, .
    \label{ae:first_moment_1}
\end{aligned}
\end{equation}
Separating the terms with \mbox{$j = i$} and those with \mbox{$j \neq i$}, we find
\begin{equation}
\begin{aligned}
    \frac{d \langle s_i \rangle}{dt} &= \sum_{\{s\}} s_i \left(E_i^{+1} - 1\right) \left[s_i r_i^- P(s_1,\ldots,s_N)\right] + \sum_{\{s\}} s_i \left(E_i^{-1} - 1\right) \left[(1 - s_i) r_i^+ P(s_1,\ldots,s_N)\right]\\[5pt]
    &+ \sum_{\{s\}} \sum_{j \neq i}^N s_i \left(E_j^{+1} - 1\right) \left[s_j r_j^- P(s_1,\ldots,s_N)\right] + \sum_{\{s\}} \sum_{j \neq i}^N s_i \left(E_j^{-1} - 1\right) \left[(1 - s_j) r_j^+ P(s_1,\ldots,s_N)\right] \, .
    \label{ae:first_moment_2}
\end{aligned}
\end{equation}
If we now use the relation~\eqref{ae:notations_relation} to extract, from the general sum over $\{s\}$, the sum over the values of $s_i$ for the terms with \mbox{$j = i$}, while we extract the sum over the values of $s_j$ for the terms with \mbox{$j \neq i$}, we obtain
\begin{equation}
\begin{aligned}
    \frac{d \langle s_i \rangle}{dt} =& \sum_{\{s\}_i} \Biggl[ \Biggl( \sum_{s_i} s_i \left( E_i^{+1} - 1 \right) \left[ s_i r_i^- P(s_1,\ldots,s_N) \right] \Biggr) + \Biggl( \sum_{s_i} s_i \left(E_i^{-1} - 1\right) \left[ (1 - s_i) r_i^+ P(s_1,\ldots,s_N)\right] \Biggr) \Biggr]\\[5pt]
    &+ \sum_{j \neq i}^N \sum_{\{s\}_j} s_i \Biggl[ \Biggl( \sum_{s_j} \left(E_j^{+1} - 1\right) \left[s_j r_j^- P(s_1,\ldots,s_N)\right] \Biggr) + \Biggl( \sum_{s_j} \left(E_j^{-1} - 1\right) \left[(1 - s_j) r_j^+ P(s_1,\ldots,s_N)\right] \Biggr) \Biggr] \, ,
    \label{ae:first_moment_3}
\end{aligned}
\end{equation}
where we can easily identify relations~\eqref{ae:relation_1} and~\eqref{ae:relation_2} for the terms with \mbox{$j \neq i$}, and relations~\eqref{ae:relation_3} and~\eqref{ae:relation_4} for the terms with \mbox{$j = i$}. In this way, we can write
\begin{equation}
    \frac{d \langle s_i \rangle}{dt} = \sum_{\{s\}_i} \left( - \sum_{s_i} s_i r_i^- P(s_1,\ldots,s_N) \right) + \sum_{\{s\}_i} \left( \sum_{s_i} (1 - s_i) r_i^+ P(s_1,\ldots,s_N) \right) \, ,
    \label{ae:first_moment_4}
\end{equation}
which, after combining the sums together again, becomes
\begin{equation}
    \frac{d \langle s_i \rangle}{dt} = \sum_{\{s\}} \left[ r_i^+ - (r_i^+ + r_i^-) s_i \right] P(s_1,\ldots,s_N) \, ,
    \label{ae:first_moment_5}
\end{equation}
and we finally find the equation for the time evolution of the first-order moments presented in the main text,
\begin{equation}
    \frac{d \langle s_i \rangle}{dt} = \langle r_i^+ \rangle - \langle (r_i^+ + r_i^-) s_i \rangle .
    \label{ae:first_moment}
\end{equation}


\subsection{Equation for the time evolution of the second-order cross-moments $\langle s_i s_j \rangle$}
\label{a:equation_for_the_time_evolution_of_the_second-order_cross-moments}

In order to find a general equation for the time evolution of the second-order cross-moments \mbox{$\langle s_i s_j \rangle$} [equation~\eqref{e:second_moment} in the main text] we proceed in a similar way as we did in the previous section for the first-order moments. Taking into account the master equation~\eqref{ae:master_equation} and using the definition of the average value in~\eqref{ae:average_definition}, we can write for the second-order cross-moments,
\begin{equation}
\begin{aligned}
    \frac{d \langle s_i s_j \rangle}{dt} &= \sum_{\{s\}} s_i s_j \frac{dP(s_1,\ldots,s_N)}{dt}\\[5pt]
    &= \sum_{\{s\}} \sum_{k=1}^N s_i s_j \left(E_k^{+1} - 1\right) \left[s_k r_k^- P(s_1,\ldots,s_N)\right] + \sum_{\{s\}} \sum_{k=1}^N s_i s_j \left(E_k^{-1} - 1\right) \left[(1 - s_k) r_k^+ P(s_1,\ldots,s_N)\right] \, .
    \label{ae:second_moment_1}
\end{aligned}
\end{equation}

For the terms of the sum with \mbox{$k \neq i, j$}, we can use relation~\eqref{ae:notations_relation} to write
\begin{equation}
    \sum_{k \neq i, j}^N \sum_{\{s\}_k} s_i s_j \Bigg[ \Bigg( \sum_{s_k} \left(E_k^{+1} - 1\right) \left[s_k r_k^- P(s_1,\ldots,s_N)\right] \Bigg) + \left( \sum_{s_k} \left(E_k^{-1} - 1\right) \left[(1 - s_k) r_k^+ P(s_1,\ldots,s_N)\right] \right) \Bigg] = 0 \, ,
    \label{ae:second_moment_2}
\end{equation}
where the equality follows from an application of relations~\eqref{ae:relation_1} and~\eqref{ae:relation_2}. Similarly, we can use relations~\eqref{ae:relation_3} and~\eqref{ae:relation_4} to transform, in equation~\eqref{ae:second_moment_1}, the terms with \mbox{$k = i \neq j$} as
\begin{equation}
\begin{aligned}
    \sum_{\{s\}_i} s_j & \Bigg[ \Bigg( \sum_{s_i} s_i \left(E_i^{+1} - 1\right) \left[s_i r_i^- P(s_1,\ldots,s_N)\right] \Bigg) + \Biggl( \sum_{s_i} s_i \left(E_i^{-1} - 1\right) \left[(1 - s_i) r_i^+ P(s_1,\ldots,s_N)\right] \Biggr) \Bigg]\\[5pt]
    &= \sum_{\{s\}_i} s_j \Biggl[ - \sum_{s_i} s_i r_i^- P(s_1,\ldots,s_N) + \sum_{s_i} (1 - s_i) r_i^+ P(s_1,\ldots,s_N) \Biggr]\\[5pt]
    &= - \sum_{\{s\}} s_i s_j r_i^- P(s_1,\ldots,s_N) + \sum_{\{s\}} (1 - s_i) s_j r_i^+ P(s_1,\ldots,s_N) \\[5pt]
    &= \langle r_i^+ s_j \rangle - \langle (r_i^+ + r_i^-) s_i s_j \rangle \, ,
    \label{ae:second_moment_3}
\end{aligned}
\end{equation}
and, equivalently, the terms with \mbox{$k = j \neq i$} as
\begin{equation}
\begin{aligned}
    \sum_{\{s\}_j} s_i & \Bigg[ \Bigg( \sum_{s_j} s_j \left(E_j^{+1} - 1\right) \left[s_j r_j^- P(s_1,\ldots,s_N)\right] \Bigg) + \Biggl( \sum_{s_j} s_j \left(E_j^{-1} - 1\right) \left[(1 - s_j) r_j^+ P(s_1,\ldots,s_N)\right] \Biggr) \Bigg]\\[5pt]
    &= \langle r_j^+ s_i \rangle - \langle (r_j^+ + r_j^-) s_i s_j \rangle \, .
    \label{ae:second_moment_4}
\end{aligned}
\end{equation}

Note that, for both expressions~\eqref{ae:second_moment_3} and~\eqref{ae:second_moment_4}, we have assumed that \mbox{$i \neq j$}. In order to study the other case, when \mbox{$i = j$}, we simply need to notice that, being the possible values of the variables \mbox{$s_i = \{0, 1\}$}, then \mbox{$s_i^2 = s_i$}, and therefore
\begin{equation}
    \frac{d \langle s_i s_i \rangle}{dt} = \frac{d \langle s_i \rangle}{dt} = \langle r_i^+ \rangle - \langle (r_i^+ + r_i^-) s_i \rangle \, ,
    \label{ae:second_moment_5}
\end{equation}
where we have used the result~\eqref{ae:first_moment} for the first-order moments derived in the previous section.

Thus, we can write an equation for the second-order cross-moments as
\begin{equation}
    \frac{d \langle s_i s_j \rangle}{dt} = \begin{cases}
    \displaystyle \langle r_i^+ s_j \rangle + \langle r_j^+ s_i \rangle - \langle q_{ij} s_i s_j \rangle & \text{if } i \neq j\\[5pt]
    \displaystyle \langle r_i^+ \rangle - \langle (r_i^+ + r_i^-) s_i \rangle & \text{if } i = j
    \end{cases} \, ,
    \label{ae:second_moment_6}
\end{equation}
where \mbox{$q_{ij} = r_i^+ + r_i^- + r_j^+ + r_j^-$}. Finally, using the Kronecker delta, we obtain the expression presented in the main text,
\begin{equation}
    \frac{d \langle s_i s_j \rangle}{dt} = \;\langle r_i^+ s_j \rangle + \langle r_j^+ s_i \rangle - \langle q_{ij} s_i s_j \rangle + \delta_{ij} \left[ \langle s_i r_i^- \rangle + \langle (1 - s_i) r_i^+ \rangle \right] \, .
    \label{ae:second_moment}
\end{equation}


\subsection{Variance of $n$}
\label{a:variance_of_n}

We derive here an analytical expression for the steady state variance of $n$ [equation~\eqref{e:variance} in the main text]. Let us start by introducing the transition rates of the noisy voter model [equation~\eqref{e:rates} in the main text] into the equation for the time evolution of the second-order cross-moments obtained in the previous section, equation~\eqref{ae:second_moment},
\begin{equation}
\begin{aligned}
    \frac{d \langle s_i s_j \rangle}{dt} =& a (\langle s_i \rangle + \langle s_j \rangle) + \frac{h}{k_i} \sum_{m \in nn(i)} \langle s_m s_j \rangle + \frac{h}{k_j} \sum_{m \in nn(j)} \langle s_m s_i \rangle - 2(2a + h) \left\langle s_i s_j \right\rangle\\[5pt]
    &+ \delta_{ij} \left[ a + h \langle s_i \rangle + \frac{h}{k_i} \sum_{m \in nn(i)} \langle s_m \rangle - \frac{2h}{k_i} \sum_{m \in nn(i)} \langle s_m s_i \rangle \right] \, .
    \label{ae:variance_1}
\end{aligned}
\end{equation}
Applying now the annealed approximation for uncorrelated networks described in the main text [see equation~\eqref{e:network_approximation}], we can replace the sums over sets of neighbors by sums over the whole system, finding
\begin{equation}
\begin{aligned}
    \frac{d \langle s_i s_j \rangle}{dt} =& a (\langle s_i \rangle + \langle s_j \rangle) + \frac{h}{N \overline{k}} \sum_m k_m \left( \langle s_m s_i \rangle + \langle s_m s_j \rangle \right) - 2(2a + h) \langle s_i s_j \rangle\\[5pt]
    &+ \delta_{ij} \Biggl[ a + h \langle s_i \rangle + \frac{h}{N \overline{k}} \sum_m k_m \langle s_m \rangle - \frac{2h}{N \overline{k}} \sum_m k_m \langle s_m s_i \rangle \Biggr] \, .
    \label{ae:variance_2}
\end{aligned}
\end{equation}

Bearing in mind the definition of the covariance matrix, \mbox{$\sigma_{ij} = \langle s_i s_j \rangle - \langle s_i \rangle \langle s_j \rangle$}, we can find an equation for its time evolution from equation~\eqref{e:kirman_first_moment} in the main text and equation~\eqref{ae:variance_2},
\begin{equation}
\begin{aligned}
    \frac{d \sigma_{ij}}{dt} =& \frac{d \langle s_i s_j \rangle}{dt} - \frac{d \langle s_i \rangle}{dt} \langle s_j \rangle - \langle s_i \rangle \frac{d \langle s_j \rangle}{dt}\\[5pt]
    =& - 2(2a + h)(\langle s_i s_j \rangle - \langle s_i \rangle \langle s_j \rangle) + \frac{h}{N \overline{k}} \sum_{m} k_m \Bigl[ (\langle s_m s_i \rangle - \langle s_m \rangle \langle s_i \rangle) + (\langle s_m s_j \rangle - \langle s_m \rangle \langle s_j \rangle) \Bigr]\\[5pt]
    &+ \delta_{ij} \left[ a + h \langle s_i \rangle + \frac{h}{N \overline{k}} \sum_{m} k_m \langle s_m \rangle - \frac{2h}{N \overline{k}} \sum_{m} k_m \langle s_m s_i \rangle \right] \, ,
    \label{ae:variance_3}
\end{aligned}
\end{equation}
which can be written in terms of only the covariance matrix and the first moments,
\begin{equation}
\begin{aligned}
    \frac{d \sigma_{ij}}{dt} =& - 2(2a + h) \sigma_{ij} + \frac{h}{N \overline{k}} \sum_{m} k_m \left( \sigma_{mi} + \sigma_{mj} \right)\\[5pt]
    &+ \delta_{ij} \left[ a + \frac{h}{N \overline{k}} \sum_{m} k_m \langle s_m \rangle + \left( h - \frac{2h}{N \overline{k}} \sum_{m} k_m \langle s_m \rangle \right) \langle s_i \rangle - \frac{2h}{N \overline{k}} \sum_{m} k_m \sigma_{mi} \right] \, .
    \label{ae:variance_4}
\end{aligned}
\end{equation}
In the steady state, and using also the steady state solution of the first order moments \mbox{$\langle s_i \rangle_{st} = 1/2$} [equation~\eqref{e:steady_state_first_moment} in the main text], we find
\begin{equation}
    \sigma_{ij} = \frac{\displaystyle\frac{h}{N \overline{k}} \sum_{m} k_m \left( \sigma_{mi} + \sigma_{mj} \right) + \delta_{ij} \left[ a + \frac{h}{2} - \frac{2h}{N \overline{k}} \sum_{m} k_m \sigma_{mi} \right]}{2(2a + h)} \, .
    \label{ae:steady_state_covariance_matrix}
\end{equation}
Note that, for the sake of notational simplicity, we have dropped the subindex $st$ for the steady state solution of the covariance matrix. Recalling now the relation between the variance of $n$ and the covariance matrix [equation~\eqref{e:variance_definition} in the main text], we can find an equation for the steady state variance of $n$ by simply summing equation~\eqref{ae:steady_state_covariance_matrix} over $i$ and $j$,
\begin{equation}
\begin{aligned}
    \sigma^2_{st} [n] &= \sum_{ij} \sigma_{ij} = \frac{\displaystyle\frac{h}{N \overline{k}} \sum_{ijm} k_m \left( \sigma_{mi} + \sigma_{mj} \right) + \sum_i \left[ a + \frac{h}{2} - \frac{2h}{N \overline{k}} \sum_{m} k_m \sigma_{mi} \right]}{2(2a + h)}\\[5pt]
    &= \frac{\displaystyle\frac{h}{\overline{k}} \left( \sum_{im} k_m \sigma_{mi} + \displaystyle\sum_{jm} k_m \sigma_{mj} \right) + N \left( a + \frac{h}{2} \right) - \frac{2h}{N \overline{k}} \sum_{im} k_m \sigma_{mi}}{2(2a + h)}\\[5pt]
    &= \frac{\displaystyle N \left( a + \frac{h}{2} \right) + \frac{2h}{\overline{k}} \left( 1 - \frac{1}{N} \right) \sum_{im} k_m \sigma_{mi}}{2(2a + h)} \, .
    \label{ae:variance_5}
\end{aligned}
\end{equation}

Let us introduce now the set of variables $S_x$, with \mbox{$x \in \{ 0,1,2,\ldots \}$}, and defined as
\begin{equation}
    S_x = \sum_{im} k_i^x k_m \sigma_{mi} \, .
    \label{ae:s_n_definition}
\end{equation}
In this way, we can rewrite the steady state variance of $n$ in terms of one of these new variables, $S_0$,
\begin{equation}
    \sigma^2_{st} [n] = \frac{\displaystyle N \left( a + \frac{h}{2} \right) + \frac{2h}{\overline{k}} \left( 1 - \frac{1}{N} \right) S_0}{2(2a + h)} \, .
    \label{ae:variance_6}
\end{equation}
In order to find an equation for this new variable $S_0$, we could use again the equation for the covariance matrix in~\eqref{ae:steady_state_covariance_matrix}, multiplying it by $k_j$ and summing over $i$ and $j$, obtaining a solution in terms of the variable $S_1$. We could then proceed similarly and find an equation for $S_1$ as a function of $S_2$, for $S_3$ as a function of $S_4$, and so forth. In general, for any $x$, we have
\begin{equation}
\begin{aligned}
    S_x &= \sum_{ij} k_i^x k_j \sigma_{ij} = \frac{\displaystyle\frac{h}{N \overline{k}} \sum_{ijm} k_i^x k_j k_m \left( \sigma_{mi} + \sigma_{mj} \right) + \sum_i k_i^{x+1} \left[ a + \frac{h}{2} - \frac{2h}{N \overline{k}} \sum_{m} k_m \sigma_{mi} \right]}{2(2a + h)}\\[5pt]
    &= \frac{\displaystyle\frac{h}{N \overline{k}} \sum_{j} k_j \sum_{im} k_i^x k_m \sigma_{mi} + \frac{h}{N \overline{k}} \sum_{i} k_i^x \sum_{jm} k_j k_m \sigma_{mj} + \sum_{i} k_i^{x+1} \left( a + \frac{h}{2} \right) - \frac{2h}{N \overline{k}} \sum_{im} k_i^{x+1} k_m \sigma_{mi}}{2(2a + h)}\\[5pt]
    &= \frac{\displaystyle h S_x + \frac{h}{\overline{k}} \overline{k^x} S_1 + N \overline{k^{x+1}} \left( a + \frac{h}{2} \right) - \frac{2h}{N \overline{k}} S_{x+1}}{2(2a + h)} \, ,
    \label{ae:variance_7}
\end{aligned}
\end{equation}
where the overbar notation is used for averages over the degree distribution [see equation~\eqref{e:overbar} in the main text]. From equation~\eqref{ae:variance_7} we can obtain an expression for the variable $S_x$ in terms of only $S_1$ and $S_{x+1}$,
\begin{equation}
    S_x = \frac{\displaystyle \frac{h}{\overline{k}} \overline{k^x} S_1 + N \overline{k^{x+1}} \left( a + \frac{h}{2} \right) - \frac{2h}{N \overline{k}} S_{x+1}}{4a + h} \, .
    \label{ae:variance_8}
\end{equation}

By inverting equation~\eqref{ae:variance_8}, we can write all variables $S_{x+1}$ in terms of the preceding ones,
\begin{equation}
    S_{x+1} = \left[ - \frac{(4a+h) N \overline{k}}{2h} \right] S_x + \frac{N}{2} \left[ \overline{k^x} S_1 + \frac{N \overline{k}}{h} \left( a + \frac{h}{2} \right) \overline{k^{x+1}} \right] \, ,
    \label{ae:variance_9}
\end{equation}
which has the general form
\begin{equation}
    S_{x+1} = A S_x + B_x \, .
    \label{ae:variance_10}
\end{equation}
It is easy to see that this recurrence relation has the solution
\begin{equation}
    S_{x+1} = A^x S_1 + \sum_{m=1}^x A^{x-m} B_x \, ,
    \label{ae:variance_11}
\end{equation}
where the choice of $S_1$ instead of $S_0$ in the first term allows us to write all the variables $S_{x+1}$ in terms of only one of them, $S_1$. Note that this choice is required by the presence of a term with $S_1$ inside $B_x$. Thus, we can write the solution for our original recurrence relation in~\eqref{ae:variance_9} as
\begin{equation}
    S_{x+1} = \left[ - \frac{(4a+h)N \overline{k}}{2h} \right]^{x} S_1 + \sum_{m=1}^{x} \left[ - \frac{(4a+h)N \overline{k}}{2h} \right]^{x-m} \frac{N}{2} \left[ \overline{k^m} S_1 + \frac{N \overline{k}}{h} \left( a + \frac{h}{2} \right) \overline{k^{m+1}} \right] \, .
    \label{ae:variance_12}
\end{equation}

If we now rewrite equation~\eqref{ae:variance_12} as
\begin{equation}
    \frac{S_{x+1}}{\displaystyle\left[ - \frac{(4a+h)N \overline{k}}{2h} \right]^{x}} = S_1 + \sum_{m=1}^{x} \left[ - \frac{(4a+h)N \overline{k}}{2h} \right]^{-m} \frac{N}{2} \left[ \overline{k^m} S_1 + \frac{N \overline{k}}{h} \left( a + \frac{h}{2} \right) \overline{k^{m+1}} \right] \, ,
    \label{ae:variance_13}
\end{equation}
we find that the left hand side of this equation vanishes in the limit of $x \to \infty$,
\begin{equation}
    \lim_{x \to \infty} \frac{S_{x+1}}{\displaystyle\left[ - \frac{(4a+h)N \overline{k}}{2h} \right]^{x}} = \lim_{x \to \infty} \frac{\displaystyle\sum_{ij} k_i^{x+1} k_j \sigma_{ij}}{\displaystyle\left[ - \frac{(4a+h)N \overline{k}}{2h} \right]^{x}} = \left[ - \frac{(4a+h)N \overline{k}}{2h} \right] \lim_{x \to \infty} \sum_{ij} \left[ - \frac{2hk_i}{(4a+h)N \overline{k}} \right]^{x+1} k_j \sigma_{ij} = 0 \, ,
    \label{ae:variance_14}
\end{equation}
where we have used the definition of the variables $S_x$ given in equation~\eqref{ae:s_n_definition}. A necessary and sufficient condition for the last equality in equation~\eqref{ae:variance_14} to hold is that
\begin{equation}
    \forall i : \left| - \frac{2hk_i}{(4a+h)N \overline{k}} \right| < 1 \; \Longrightarrow \; \forall i : k_i < \frac{(4a+h)N\overline{k}}{2h} \, ,
    \label{ae:condition}
\end{equation}
which is generally true and always true for $h>0$ and $\overline{k} \geq 2$. Thus, in the $x \to \infty$ limit, we can equate the right hand side of equation~\eqref{ae:variance_13} to zero,
\begin{equation}
    S_1 + \left( \sum_{m=1}^{\infty} \left[ - \frac{(4a+h)N \overline{k}}{2h} \right]^{-m} \frac{N}{2} \overline{k^m} \right) S_1 + \left( \sum_{m=1}^{\infty} \left[ - \frac{(4a+h)N \overline{k}}{2h} \right]^{-m} \frac{N^2 \overline{k}}{2h} \left( a + \frac{h}{2} \right) \overline{k^{m+1}} \right) = 0 \, ,
    \label{ae:variance_15}
\end{equation}
and find, in this way, a solution for $S_1$,
\begin{equation}
    S_1 = \frac{- \displaystyle\frac{N^2 \overline{k}}{2h} \left( a + \frac{h}{2} \right) \sum_{m=1}^{\infty} \left[ \frac{-2h}{(4a+h)N \overline{k}} \right]^m \overline{k^{m+1}}}{1 + \displaystyle\frac{N}{2} \sum_{m=1}^{\infty} \left[ \frac{-2h}{(4a+h)N \overline{k}} \right]^m \overline{k^m}} \, .
    \label{ae:variance_16}
\end{equation}

Regarding the sums in equation~\eqref{ae:variance_16}, we can use the sum of the geometric series
\begin{equation}
    \sum_{m=1}^{\infty} A^{m} \overline{k^{m+z}} = \overline{k^z \sum_{m=1}^{\infty} A^{m} k^{m}} = \overline{\frac{A k^{z+1}}{1-Ak}} \, , \quad \textrm{if} \quad |Ak|< 1 \, ,
    \label{ae:variance_17}
\end{equation}
where the condition of convergence is exactly the same as presented before in equation~\eqref{ae:condition}, and thus generally true and always true for $h>0$ and $\overline{k} \geq 2$. In this way, applying the result~\eqref{ae:variance_17} to equation~\eqref{ae:variance_16} we have
\begin{equation}
    S_1 = \frac{N^2 \overline{k} \displaystyle\left( a + \frac{h}{2} \right) \overline{\left( \frac{k^2}{1 + \frac{2hk}{(4a+h)N \overline{k}}} \right)}}{(4a+h) N \overline{k} - h N \overline{\displaystyle\left( \frac{k}{1 + \frac{2hk}{(4a+h)N\overline{k}}} \right)}} = \frac{N^2 \overline{k} \displaystyle\left( a + \frac{h}{2} \right) (4a+h) \overline{\left( \frac{k^2}{(4a+h)N \overline{k} + 2hk} \right)}}{4a + h - \displaystyle\frac{h}{\overline{k}} \overline{\left( \frac{(4a+h) N \overline{k} k}{(4a+h) N \overline{k} + 2hk} \right)}} \, ,
    \label{ae:variance_18}
\end{equation}
where the denominator can be rewritten as
\begin{equation}
\begin{aligned}
    4a + h - \displaystyle\frac{h}{\overline{k}} \overline{\left( \frac{(4a+h) N \overline{k} k}{(4a+h) N \overline{k} + 2hk} \right)} &= 4a + \frac{h}{\overline{k}} \overline{\frac{[(4a+h) N \overline{k} + 2hk] \overline{k} - (4a+h) N \overline{k} k}{(4a+h) N \overline{k} + 2hk}}\\[5pt]
    &= 4a + \frac{h}{\overline{k}} \overline{\frac{[(4a+h) N \overline{k} + 2hk] (\overline{k} - k) + 2hk^2}{(4a+h) N \overline{k} + 2hk}}\\[5pt]
    &= 4a + \frac{h}{\overline{k}} \overline{(\overline{k} - k)} + \frac{2h^2}{\overline{k}} \overline{\left(\frac{k^2}{(4a+h) N \overline{k} + 2hk}\right)}\\[5pt]
    &= 4a + \frac{2h^2}{\overline{k}} \overline{\left(\frac{k^2}{(4a+h) N \overline{k} + 2hk}\right)} \, ,
    \label{ae:variance_19}
\end{aligned}
\end{equation}
thereby finding a final expression for $S_1$,
\begin{equation}
    S_1 = \frac{N^2 \overline{k} \displaystyle\left( a + \frac{h}{2} \right) (4a+h) \overline{\left( \frac{k^2}{(4a+h)N \overline{k} + 2hk} \right)}}{4a + \displaystyle\frac{2h^2}{\overline{k}} \overline{\left(\frac{k^2}{(4a+h) N \overline{k} + 2hk}\right)}} \, .
    \label{ae:s_1_final}
\end{equation}

If we now go back to the equation for the steady state variance $\sigma^2_{st}[n]$ as a function of $S_0$, equation~\eqref{ae:variance_6}, and we use equation~\eqref{ae:variance_8} to find an expression for $S_0$ as a function of $S_1$,
\begin{equation}
    S_0 = \frac{\displaystyle N \overline{k} \left( a + \frac{h}{2} \right) + \frac{h}{\overline{k}} \left( 1 - \frac{2}{N} \right) S_1}{4a + h} \, ,
    \label{ae:variance_20}
\end{equation}
then we can write an equation for the steady state variance as a function of $S_1$,
\begin{equation}
    \sigma^2_{st} [n] = \frac{N}{4} \left[ 1 + \frac{2h\left( 1 - \frac{1}{N} \right)}{4a + h} + \left( N - 3 + \frac{2}{N} \right) \left( \frac{h}{\overline{k}} \right)^2 \frac{2 S_1}{N^2 \left( a + \frac{h}{2} \right) (4a+h)} \right] \, .
    \label{ae:variance_21}
\end{equation}
Finally, introducing here what we found for $S_1$ in equation~\eqref{ae:s_1_final}, we arrive to the final expression for the steady state variance of the global variable $n$ as presented in the main text,
\begin{equation}
    \sigma^2_{st} [n] = \frac{N}{4} \left[ 1 + \frac{2h\left( 1 - \frac{1}{N} \right)}{4a + h} + \left( N - 3 + \frac{2}{N} \right) \frac{\displaystyle\left(\frac{h^2}{\overline{k}}\right) \overline{\left( \frac{k^2}{(4a+h)N\overline{k} + 2hk} \right)}}{2a + \displaystyle\left(\frac{h^2}{\overline{k}}\right) \overline{\left(\frac{k^2}{(4a+h) N \overline{k} + 2hk}\right)}} \right] \, .
    \label{ae:variance}
\end{equation}


\subsection{Asymptotic approximations for the variance of $n$}
\label{a:asymptotic_approximations_for_the_variance_of_n}

We develop here a first-order approximation for the steady state variance of $n$ with respect to the system size $N$. Given the dependence of the result of this approximation on the relationship between the system size $N$ and the noise parameter $a$, we are forced to consider two different asymptotic approximation regimes: one for small $a$ [corresponding to equation~\eqref{e:variance_small_a} in the main text] and the other for large $a$ [corresponding to equation~\eqref{e:variance_large_a} in the main text].

Let us start by noticing that the structural constraint imposed by the annealed approximation for uncorrelated networks on the degrees of the network, $k_i < \sqrt{N \overline{k}}$, allows us to write equation~\eqref{ae:variance} as
\begin{equation}
    \sigma^2_{st} [n] = \frac{N}{4} \left[ 1 + \frac{2h\left( 1 - \frac{1}{N} \right)}{4a + h} + \left( N - 3 + \frac{2}{N} \right) \frac{\displaystyle\left(\frac{h^2}{\overline{k}}\right) \overline{\left( \frac{k^2}{(4a+h)N\overline{k} \left( 1 + \mathcal{O}\left( N^{-1/2} \right) \right)} \right)}}{2a + \displaystyle\left(\frac{h^2}{\overline{k}}\right) \overline{\left(\frac{k^2}{(4a+h) N \overline{k} \left( 1 + \mathcal{O}\left( N^{-1/2} \right) \right)}\right)}} \right] \, .
    \label{ae:approx_1}
\end{equation}
In this way, we notice that, depending on the order of the product $aN$, the approximation of the third term in equation~\eqref{ae:approx_1} will lead to different results. In particular, when the noise parameter $a$ is of order $\mathcal{O}(N^{-1})$ or smaller, then the product $aN$ is, at most, of order $\mathcal{O}(N^0)$, and we can continue with the approximation as
\begin{equation}
\begin{aligned}
    \sigma^2_{st} [n] &= \frac{N}{4} \left[ 1 + \frac{2h\left( 1 - \frac{1}{N} \right)}{4a + h} + \left( N - 3 + \frac{2}{N} \right) \left( \frac{\displaystyle\left(\frac{h^2}{\overline{k}}\right) \left( \frac{\overline{k^2}}{(4a+h)N \overline{k}} \right)}{2a + \displaystyle\left(\frac{h^2}{\overline{k}}\right) \left(\frac{\overline{k^2}}{(4a+h) N \overline{k}}\right)} + \mathcal{O}(N^{-1/2}) \right) \right]\\[5pt]
     &= \frac{N}{4} \left[ 1 + 2\left( 1 - \frac{1}{N} \right) + \left( N - 3 + \frac{2}{N} \right) \left( \frac{h \displaystyle\left(\frac{\overline{k^2}}{\overline{k}^2}\right)}{2aN + h \displaystyle\left(\frac{\overline{k^2}}{\overline{k}^2}\right)} + \mathcal{O}(N^{-1/2}) \right) \right] \, ,
    \label{ae:approx_2}
\end{aligned}
\end{equation}
which, to the first order in $N$, becomes
\begin{equation}
    \sigma^2_{st} [n] = \frac{N}{4} \left[ N \left( \frac{h \displaystyle\left(\frac{\overline{k^2}}{\overline{k}^2}\right)}{2aN + h \displaystyle\left(\frac{\overline{k^2}}{\overline{k}^2}\right)} + \mathcal{O}(N^{-1/2}) \right) \right] \, .
    \label{ae:approx_3}
\end{equation}
Using now the definition of the variance of the degree distribution, \mbox{$\sigma^2_k = \overline{k^2} - \overline{k}^2$}, we find the approximation presented in the main text for the steady state variance of $n$ for small $a$ and to the first order in $N$,
\begin{equation}
    \sigma^2_{st}[n] = \frac{N^2}{4} \left[ \frac{ \displaystyle h \left( \frac{\sigma^2_k}{\overline{k}^2} + 1 \right)}{\displaystyle 2aN + h \left( \frac{\sigma^2_k}{\overline{k}^2} + 1 \right)} \right] + \mathcal{O}(N^{3/2}) \, .
    \label{ae:variance_small_a}
\end{equation}
Note that the remaining terms are \emph{at most} of order $\mathcal{O}(N^{3/2})$.

On the contrary, when $a$ is of order $\mathcal{O}(N^0)$ or larger, then the product $aN$ is, at least, of order $\mathcal{O}(N)$, and we can approximate equation~\eqref{ae:approx_1} as
\begin{equation}
\begin{aligned}
    \sigma^2_{st} [n] &= \frac{N}{4} \left[ 1 + \frac{2h\left( 1 - \frac{1}{N} \right)}{4a + h} + \left( N - 3 + \frac{2}{N} \right) \left( \frac{\displaystyle\left(\frac{h^2}{\overline{k}}\right) \left( \frac{\overline{k^2}}{(4a+h)N \overline{k}} \right)}{2a + \displaystyle\left(\frac{h^2}{\overline{k}}\right) \left(\frac{\overline{k^2}}{(4a+h) N \overline{k}}\right)} + \mathcal{O}(N^{-3/2}) \right) \right]\\[5pt]
     &= \frac{N}{4} \left[ 1 + \frac{2h\left( 1 - \frac{1}{N} \right)}{4a + h} + \left( N - 3 + \frac{2}{N} \right) \left( \frac{h^2 \displaystyle\left(\frac{\overline{k^2}}{\overline{k}^2}\right)}{2a(4a+h)N + h^2 \displaystyle\left(\frac{ \overline{k^2}}{\overline{k}^2}\right)} + \mathcal{O}(N^{-3/2}) \right) \right]\\[5pt]
     &= \frac{N}{4} \left[ 1 + \frac{2h\left( 1 - \frac{1}{N} \right)}{4a + h} + \left( N - 3 + \frac{2}{N} \right) \left( \frac{h^2 \displaystyle\left(\frac{\overline{k^2}}{\overline{k}^2}\right)}{2a(4a+h)N} + \mathcal{O}(N^{-3/2}) \right) \right] \, .
    \label{ae:approx_4}
\end{aligned}
\end{equation}
Note that the remaining terms are now one order of $N$ smaller than in the previous approximation [equation~\eqref{ae:approx_2}]. To the first order in $N$ we have
\begin{equation}
    \sigma^2_{st} [n] = \frac{N}{4} \left[ 1 + \frac{2h}{4a + h} + \frac{h^2 \displaystyle\left(\frac{\overline{k^2}}{\overline{k}^2}\right)}{2a(4a+h)} + \mathcal{O}(N^{-1/2}) \right] = \frac{N}{4} \left[ 1 + \frac{\displaystyle 4ah + h^2 \left( \frac{\overline{k^2} - \overline{k}^2 + \overline{k}^2}{\overline{k}^2} \right) }{2a(4a+h)} + \mathcal{O}(N^{-1/2}) \right] \, ,
    \label{ae:approx_5}
\end{equation}
and, finally, we find the approximation presented in the main text for the steady state variance of $n$ for large $a$ and to the first order in $N$,
\begin{equation}
    \sigma^2_{st}[n] = \frac{N}{4} \left[ 1 + \frac{h}{2a} + \frac{\displaystyle h^2 \frac{\sigma^2_k}{\overline{k}^2} }{2a ( 4a + h)} \right] + \mathcal{O}(N^{1/2}) \, ,
    \label{ae:variance_large_a}
\end{equation}
where the remaining terms are \emph{at most} of order $\mathcal{O}(N^{1/2})$.


\subsection{Critical point approximation}
\label{a:critical_point_approximation}

In this section, we derive an analytical approximation for the critical point of the bimodal-unimodal transition [equation~\eqref{e:critical_point} in the main text], which can be defined as the relationship between the model parameters $a$ and $h$ leading the steady state variance of $n$ to take the value \mbox{$\sigma_{st}^2[n] = N(N+2)/12$}, corresponding to a uniform distribution between $0$ and $N$. In particular, bearing in mind that the critical value $a_c$ of a fully-connected system is of order $\mathcal{O}(N^{-1})$ and that the change due to the network structure appears to be of order $\mathcal{O}(N^0)$ (see Fig.~\ref{f:varianceExact} in the main text), then we can expect the value of the critical point to be still of order $\mathcal{O}(N^{-1})$, and we can therefore use the small~$a$ asymptotic approximation in equation~\eqref{ae:variance_small_a},
\begin{equation}
    \sigma^2_{st}[n] = \frac{N^2}{4} \left[ \frac{ \displaystyle h \left( \frac{\sigma^2_k}{\overline{k}^2} + 1 \right)}{\displaystyle 2a_c N + h \left( \frac{\sigma^2_k}{\overline{k}^2} + 1 \right)} \right] + \mathcal{O}(N^{3/2}) = \frac{N(N+2)}{12} \, .
    \label{ae:critical_point_1}
\end{equation}
The solution of this equation leads to the, for large $N$, leads to the value of the critical point discussed in the main text,
\begin{equation}
    a_c = \frac{h}{N} \left( \frac{\sigma^2_k}{\overline{k}^2} + 1 \right) + \mathcal{O}(N^{-3/2}) \, ,
    \label{ae:critical_point}
\end{equation}
consistent with the assumption of a critical value of order $\mathcal{O}(N^{-1})$. Note that assuming, instead, the critical value to be of order $\mathcal{O}(N^{0})$, and using therefore the large~$a$ asymptotic approximation in equation~\eqref{ae:variance_large_a}, leads again to an $a_c$ of order $\mathcal{O}(N^{-1})$, inconsistent with the initial assumption.


\subsection{Order parameter: the interface density $\rho$}
\label{a:order_parameter_the_interface_density}

We obtain, this section, an analytical expression for the order parameter $\rho$ [equation~\eqref{e:rho} in the main text]. $\rho$ is defined as the interface density or density of active links, that is, the fraction of links connecting nodes in different states. In terms of the connectivity matrix $A_{ij}$,
\begin{equation}
    \rho = \frac{\displaystyle\frac{1}{2} \sum_{ij} A_{ij} [s_i(1-s_j) + (1-s_i)s_j]}{\displaystyle\frac{1}{2} \sum_{ij} A_{ij}} = \frac{\displaystyle \sum_{ij} A_{ij} (s_i + s_j - s_i s_j)}{\displaystyle \sum_{ij} A_{ij}} \, ,
    \label{ae:rho_definition}
\end{equation}
and introducing the annealed approximation for uncorrelated networks described in the main text [see equation~\eqref{e:network_approximation}], we find
\begin{equation}
    \rho = \frac{\displaystyle \sum_{ij} \frac{k_i k_j}{N \overline{k}} (s_i + s_j - s_i s_j)}{\displaystyle \sum_{ij} \frac{k_i k_j}{N \overline{k}}} = \sum_{ij} \frac{k_i k_j}{\left( N \overline{k} \right)^2} (s_i + s_j - s_i s_j) \, .
    \label{ae:rho_1}
\end{equation}
Restricting our attention to the steady state average value of equation~\eqref{ae:rho_1},
\begin{equation}
    \langle \rho \rangle_{st} = \sum_{ij} \frac{k_i k_j}{\left( N \overline{k} \right)^2} \bigl( \langle s_i \rangle_{st} + \langle s_j \rangle_{st} - \langle s_i s_j \rangle_{st} \bigr) \, ,
    \label{ae:rho_2}
\end{equation}
we can use the steady state mean solution found before for the individual node variables $s_i$, \mbox{$\langle s_i \rangle_{st} = 1/2$}, and the definition of the covariance matrix in the steady state, \mbox{$\sigma_{ij} = \langle s_i s_j \rangle_{st} - 1/4$}, in order to write
\begin{equation}
    \langle \rho \rangle_{st} = \frac{1}{2} - \frac{2}{\left( N \overline{k} \right)^2} \sum_{ij} k_i k_j \sigma_{ij} \, ,
    \label{ae:rho_3}
\end{equation}
where we can identify the variable $S_1$ [see equation~\eqref{ae:s_n_definition}],
\begin{equation}
    \langle \rho \rangle_{st} = \frac{1}{2} - \frac{2 S_1}{\left( N \overline{k} \right)^2}.
    \label{ae:rho_4}
\end{equation}
Finally, reversing the relation~\eqref{ae:variance_21} between the variance of $n$ and the variable $S_1$, we can write the steady state average interface density $\rho$ in terms of the variance of $n$,
\begin{equation}
    \langle \rho \rangle_{st} = \frac{1}{2} - \frac{2}{(hN)^2} \left[ \frac{(4a+h)(2a+h)}{\left( 1 - \frac{1}{N} \right) \left( 1 - \frac{2}{N} \right)} \left( \sigma^2[n] - \frac{N}{4} \right) - \frac{\left( a + \frac{h}{2} \right)}{\left( 1 - \frac{2}{N} \right)} hN \right] \, ,
    \label{ae:rho_5}
\end{equation}
as it appears in the main text.


\subsection{Autocorrelation function of $n$}
\label{a:autocorrelation_function_of_n}

We derive here an analytical expression for the steady state autocorrelation function of $n$ [equations~\eqref{e:autocorrelation} and~\eqref{e:autocorrelation_C} in the main text], defined as
\begin{equation}
    K_{st}[n](\tau) = \langle n(t + \tau) n(t) \rangle_{st} - \langle n \rangle_{st}^2 \, ,
    \label{ae:autocorrelation_definition}
\end{equation}
where $\tau$ plays the role of a time-lag. As far as the second point in time, $t + \tau$, is concerned, we assume that the system was at $n(t)$ at time $t$, and hence we can treat $n(t)$ as an initial condition,
\begin{equation}
    K_{st}[n](\tau) = \langle \langle n(t + \tau) | n(t) \rangle n(t) \rangle_{st} - \langle n \rangle_{st}^2 \, ,
    \label{ae:autocorrelation_1}
\end{equation}
which, in terms of the individual variables $\{s_i\}$ and taking into account that $\langle n \rangle_{st} = N/2$, can be written as
\begin{equation}
    K_{st}[n](\tau) = \sum_{ij} \langle \langle s_i(t + \tau) | \{s_l(t)\} \rangle s_j(t) \rangle_{st} - \frac{N^2}{4} \, .
    \label{ae:autocorrelation_2}
\end{equation}
We need, therefore, an expression for \mbox{$\langle s_i(t + \tau) | \{s_l(t)\} \rangle$}, which we find by integration of the equation for the temporal evolution of the first-order moments $\langle s_i \rangle$ ---obtained by introducing the transition rates of the noisy voter model into equation~\eqref{ae:first_moment}---,
\begin{equation}
    \frac{d \langle s_i(t + \tau) | \{s_l(t)\} \rangle}{d\tau} = a - (2a + h) \langle s_i(t + \tau) | \{s_l(t)\} \rangle + \frac{h}{N \overline{k}} \sum_m k_m \langle s_m(t + \tau) | \{s_l(t)\} \rangle \, .
    \label{ae:autocorrelation_3}
\end{equation}

In order to integrate equation~\eqref{ae:autocorrelation_3}, we must first obtain an expression for
\begin{equation}
    b(t + \tau) \equiv \frac{h}{N \overline{k}} \sum_m k_m \langle s_m(t + \tau) | \{s_l(t)\} \rangle \, ,
    \label{ae:definition_b}
\end{equation}
which we can find by multiplying equation~\eqref{ae:autocorrelation_3} by $hk_i/N\overline{k}$ and summing over $i$, 
\begin{equation}
\begin{aligned}
    \frac{d}{d\tau} \left( \frac{h}{N \overline{k}} \sum_i k_i \langle s_i(t + \tau) | \{s_l(t)\} \rangle\right) =& \frac{ah}{N\overline{k}} \sum_i k_i - \frac{(2a + h)h}{N \overline{k}} \sum_i k_i \langle s_i(t + \tau) | \{s_l(t)\} \rangle\\[5pt]
    &+ \left( \frac{h}{N \overline{k}} \right)^2 \sum_i k_i \sum_m k_m \langle s_m(t + \tau) | \{s_l(t)\} \rangle \, .
    \label{ae:autocorrelation_4}
\end{aligned}
\end{equation}
In this way, we arrive to the differential equation
\begin{equation}
    \frac{db(t+\tau)}{d\tau} = ah - (2a+h)b(t+\tau) +hb(t+\tau) = ah - 2ab(t+\tau) \, ,
    \label{ae:autocorrelation_5}
\end{equation}
which has the solution
\begin{equation}
    b(t + \tau) = \frac{h}{2} \left( 1 - e^{-2a\tau} \right) + b(t) e^{-2a\tau} \, ,
    \label{ae:autocorrelation_6}
\end{equation}
depending on the initial condition $b(t)$. Using this expression, we can now integrate equation~\eqref{ae:autocorrelation_3} for the first-order moments,
\begin{equation}
    \frac{d \langle s_i(t + \tau) | \{s_l(t)\} \rangle}{d\tau} = a - (2a + h) \langle s_i(t + \tau) | \{s_l(t)\} \rangle + b(t + \tau) \, ,
    \label{ae:autocorrelation_7}
\end{equation}
which has the general solution
\begin{equation}
\begin{aligned}
    \langle s_i(t + \tau) | \{s_l(t)\} \rangle &= \frac{\displaystyle \int_{0}^{\tau} e^{(2a+h)\tau'} \left[ a + b(t + \tau') \right] d\tau' + c_1}{e^{(2a+h)\tau}}\\[5pt]
    &= \frac{\displaystyle \int_{0}^{\tau} e^{(2a+h)\tau'} \left[ a + \frac{h}{2} \left( 1 - e^{-2a\tau'} \right) + b(t) e^{-2a\tau'} \right] d\tau' + c_1}{e^{(2a+h)\tau}}\\[5pt]
    &= \frac{\displaystyle \left( a + \frac{h}{2} \right) \int_{0}^{\tau} e^{(2a+h)\tau'} d\tau' + \left( b(t) - \frac{h}{2} \right) \int_{0}^{\tau} e^{h\tau'} d\tau' + c_1}{e^{(2a+h)\tau}}\\[5pt]
    &= \frac{1}{2} \left( 1 - e^{-(2a+h)\tau} \right) + \frac{b(t) - \frac{h}{2}}{h} \left( e^{-2a\tau} - e^{-(2a+h)\tau} \right) + c_1 e^{-(2a+h)\tau} \, .
    \label{ae:autocorrelation_8}
\end{aligned}
\end{equation}
Applying now the initial condition \mbox{$\langle s_i(t) | \{s_l(t)\} \rangle = s_i(t)$}, we find
\begin{equation}
    \langle s_i(t + \tau) | \{s_l(t)\} \rangle = \frac{1}{2} \left( 1 - e^{-(2a+h)\tau} \right) + \frac{b(t) - \frac{h}{2}}{h} \left( e^{-2a\tau} - e^{-(2a+h)\tau} \right) + s_i(t) e^{-(2a+h)\tau} \, .
    \label{ae:autocorrelation_9}
\end{equation}

We are now ready to go back to the autocorrelation function~\eqref{ae:autocorrelation_2} and write, in the steady state,
\begin{equation}
\begin{aligned}
    K_{st}[n](\tau) =& \sum_{ij} \left\langle \frac{1}{2} \left( 1 - e^{-(2a+h)\tau} \right) s_j(t) \right\rangle_{st} + \sum_{ij} \left\langle \frac{b(t) - \frac{h}{2}}{h} \left( e^{-2a\tau} - e^{-(2a+h)\tau} \right) s_j(t) \right\rangle_{st}\\[5pt]
    &+ \sum_{ij} \left\langle s_i(t) s_j(t) e^{-(2a+h)\tau} \right\rangle_{st} - \frac{N^2}{4} \, .
    \label{ae:autocorrelation_10}
\end{aligned}
\end{equation}
Given that we assume the state of the system at $t$ to be our initial condition, $b(t)$ can be written as
\begin{equation}
    b(t) = \frac{h}{N \overline{k}} \sum_i k_i \langle s_i(t) | \{s_l(t)\} \rangle = \frac{h}{N \overline{k}} \sum_i k_i s_i(t) \, ,
    \label{ae:autocorrelation_11}
\end{equation}
and thus we have, for the autocorrelation function,
\begin{equation}
\begin{aligned}
    K_{st}[n](\tau) =& \frac{1}{2} \left( 1 - e^{-(2a+h)\tau} \right) \sum_{ij} \langle s_j(t) \rangle_{st} + \frac{1}{N\overline{k}} \left( e^{-2a\tau} - e^{-(2a+h)\tau} \right) \sum_{ijm} k_m \langle s_m(t) s_j(t) \rangle_{st}\\[5pt]
    &- \frac{1}{2} \left( e^{-2a\tau} - e^{-(2a+h)\tau} \right) \sum_{ij} \langle s_j(t) \rangle_{st} + e^{-(2a+h)\tau} \sum_{ij} \langle s_i(t) s_j(t) \rangle_{st} - \frac{N^2}{4} \, .
    \label{ae:autocorrelation_12}
\end{aligned}
\end{equation}
Using now the value found before for the steady state solution of the first-order moments, \mbox{$\langle s_i \rangle_{st} = 1/2$}, and the definition of the covariance matrix in the steady state, \mbox{$\sigma_{ij} = \langle s_i s_j \rangle_{st} - \langle s_i \rangle_{st}^2 = \langle s_i s_j \rangle_{st} - 1/4$}, we find
\begin{equation}
    K_{st}[n](\tau) = -e^{-2a\tau} \frac{N^2}{4} + \frac{1}{\overline{k}} \left( e^{-2a\tau} - e^{-(2a+h)\tau} \right) \sum_{jm} k_m \left( \sigma_{mj} + \frac{1}{4} \right) + e^{-(2a+h)\tau} \sum_{ij} \left( \sigma_{ij} + \frac{1}{4} \right) \, .
    \label{ae:autocorrelation_13}
\end{equation}
Finally, identifying in the previous equation the variance of $n$ and the variable $S_1$ [see equation~\eqref{ae:s_n_definition}], and reordering terms according to their exponential decay, we find the expression for the autocorrelation function of $n$ discussed in the main text,
\begin{equation}
    K_{st}[n](\tau) = \left( \sigma^2[n] - \frac{S_1}{\overline{k}} \right) e^{-(2a+h)\tau} + \frac{S_1}{\overline{k}} e^{-2a\tau} \, .
    \label{ae:autocorrelation}
\end{equation}
The definition of the variable $S_1$ given in the main text, as a function of the variance of $n$, can be directly obtained by reversing equation~\eqref{ae:variance_21}.


\subsection{Suppementary Figure S1}
\label{a:suppementary_Figure_S1}

\renewcommand\thefigure{S\arabic{figure}}

    \begin{figure}[ht!]
    \centering
    \includegraphics[width=14.5cm, height=!]{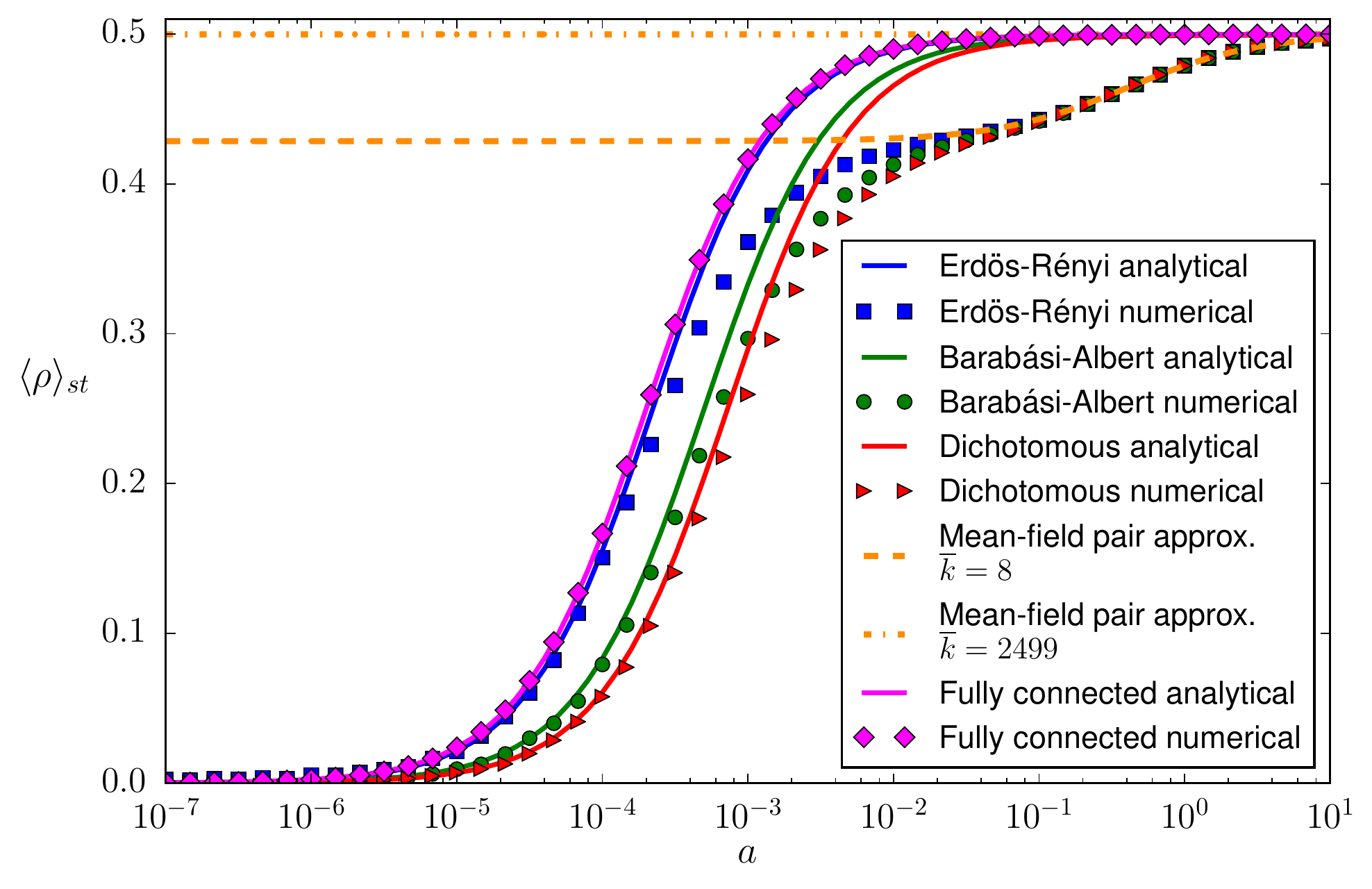}
    \caption{\label{f:rhoExactLogXSUPPL} Steady state of the average interface density as a function of the noise parameter $a$ in a linear-logarithmic scale and for three different types of networks with mean degree $\overline{k} = 8$: Erd\"os-R\'enyi random network, Barab\'asi-Albert scale-free network and dichotomous network. A fully connected topology is also included for comparisson. Symbols: Numerical results (averages over $20$ networks, $10$ realizations per network and $50000$ time steps per realization). Solid lines: Analytical results [see equation~\eqref{ae:rho_5}]. Dashed line: Mean-field pair-approximation (see \cite{Diakonova2015}) for a mean degree $\overline{k} = 8$. Dash-dotted line: Mean-field pair-approximation for a mean degree $\overline{k} = 2499$. The interaction parameter is fixed as $h = 1$ and the system size as $N = 2500$.}
    \end{figure}

\end{appendices}




\section*{Acknowledgements}

We are particularly grateful to Luis F. Lafuerza for helpful suggestions during the early stages of this work. We acknowledge financial support by FEDER (EU) and MINECO (Spain) under Grant ESOTECOS (FIS2015-63628-CZ-Z-R). AC acknowledges support by the FPU program of MECD.

\section*{Author contributions statement}

A.C., R.T. and M.S.M. conceived and designed the research; A.C. performed the numerical simulations and analyzed the results; A.C., R.T. and M.S.M. wrote and revised the manuscript.

\end{document}